\documentclass[amsmath,amssymb,aps,floats,prb,twocolumn,showpacs,superscriptaddress,longbibliography]{revtex4-2}
\usepackage{float}
\usepackage[english]{babel}
\usepackage[final]{graphicx}
\usepackage{amssymb,amsmath}
\usepackage{color}
\usepackage{xcolor}
\usepackage{longtable}
\usepackage{placeins}
\usepackage{graphicx}
\usepackage{epstopdf}
\usepackage[normalem]{ulem}
\usepackage{dcolumn}
\usepackage{bm}
\usepackage{tabularx}
\usepackage[subnum]{cases}
\usepackage[dvipsnames]{xcolor}
\usepackage{soul}

\begin{document}

\title{Electronic excitations in the Shastry-Sutherland compound SrCu$_2$(BO$_3$)$_2$}

\author{Tariq Leinen}
\affiliation{Institute of Applied Physics, University of Bern, CH-3012 Bern, Switzerland}
\author{Ola K. Forslund}
\affiliation{Institute of Physics, University of Zurich, CH-8057 Zürich, Switzerland}
\affiliation{Department of Physics and Astronomy, Uppsala University, Box 516, SE-75120 Uppsala, Sweden}
\author{Eugenio Paris}
\affiliation{PSI Center for Photon Science, Paul Scherrer Institute, CH-5232 Villigen-PSI, Switzerland}%
\author{Nicola Colonna}
\affiliation{PSI Center for Photon Science, Paul Scherrer Institute, CH-5232 Villigen-PSI, Switzerland}%
\author{Marco Caputo}
\affiliation{MAX IV Laboratory, Lund University, SE-221 00 Lund, Sweden}
\author{Johan Chang}
\affiliation{Institute of Physics, University of Zurich, CH-8057 Zürich, Switzerland}
\author{Gabriel Nagamine}
\affiliation{Institute of Applied Physics, University of Bern, CH-3012 Bern, Switzerland}
\author{Takashi Tokushima}
\affiliation{MAX IV Laboratory, Lund University, SE-221 00 Lund, Sweden}
\author{Conny Såthe}
\affiliation{MAX IV Laboratory, Lund University, SE-221 00 Lund, Sweden}
\author{Pascal Puphal}
\affiliation{Max Planck Institute for Solid State Research, Heisenbergstrasse 1, 70569 Stuttgart, Germany}%
\author{Jeremie Teyssier}
\affiliation{Department of Quantum Matter Physics, University of Geneva, 1211, Geneva, Switzerland}
\author{Thorsten Schmitt}
\affiliation{PSI Center for Photon Science, Paul Scherrer Institute, CH-5232 Villigen-PSI, Switzerland}%
\author{Nikolay A. Bogdanov}
\affiliation{Max Planck Institute for Solid State Research, Heisenbergstrasse 1, 70569 Stuttgart, Germany}%
\author{Maria Daghofer}
\affiliation{Institute for functional matter and Quantum Technologies, Universität Stuttgart, 70550 Stuttgart, Germany}
\author{Adrian L. Cavalieri}
\affiliation{Institute of Applied Physics, University of Bern, CH-3012 Bern, Switzerland}
\affiliation{PSI Center for Photon Science, Paul Scherrer Institute, CH-5232 Villigen-PSI, Switzerland}%
\author{Flavio Giorgianni}
\email{flavio.giorgianni@unibe.ch}
\affiliation{Institute of Applied Physics, University of Bern, CH-3012 Bern, Switzerland}


\begin{abstract}
SrCu$_2$(BO$_3$)$_2$ (SCBO) is a paradigmatic realization of the Shastry–Sutherland model, hosting geometrically frustrated spin dimers and a variety of quantum magnetic phases and phenomena. Although its magnetic properties have been extensively studied, the high-energy electronic excitations that determine the crystal-field environment and Cu–O hybridization have remained largely unexplored. Here we combine Cu $L_3$-edge resonant inelastic x-ray scattering (RIXS), broadband optical spectroscopy, and electronic-structure calculations to determine the relevant local and interband excitation energy scales in SCBO. RIXS resolves a well-defined manifold of localized Cu$^{2+}$ $d$--$d$ excitations between 1.8 and 2.4 eV, whose energies and polarization dependence are well reproduced by multireference quantum-chemistry calculations. In contrast, optical spectroscopy identifies charge-transfer excitations with an absorption onset near 1.2--1.6 eV and a broader higher-energy structure around 4.5 eV, which are qualitatively captured by DFT+\(U\) calculations. Taken together, these results define the characteristic energy scales of $d$–$d$ and CT excitations, offering quantitative benchmarks for computational frameworks and providing essential input for refining superexchange-based magnetic models of this prototypical frustrated quantum antiferromagnet.
\end{abstract}

\maketitle


\section{Introduction}
Quantum magnetic materials, which are characterized by strong electronic correlations, geometric frustration, and reduced dimensionality, are ideal systems for the investigation of many-body phenomena, including quantum spin liquids, fractionalization, quantum phase transitions, and topological excitations \cite{balents_spin_2010, vasiliev_milestones_2018, ramirez_short-range_2025, Bose_spingaps_2005, lee_end_2008}. Among these materials, SrCu$_2$(BO$_3$)$_2$ (SCBO) serves as a prototypical system for studying how frustration and quantum criticality manifest in two-dimensional spin networks~\cite{kageyama_exact_1999}. In SCBO, Cu$^{2+}$ ions carrying spin $S=1/2$ moments form a two-dimensional lattice of orthogonal spin dimers, realizing a perfectly geometrically frustrated Shastry--Sutherland model~\cite{shastry_exact_1981}. This unique geometry gives rise to an extraordinarily rich phase diagram, featuring fractional magnetization plateaus, quantum critical points, and exotic phases such as spin plaquettes and supersolids \cite{kageyama_anomalous_1999, zayed_4-spin_2017, nomura_unveiling_2023, shi_discovery_2022, miyahara_theory_2003, kodama_magnetic_2002, laflorencie_quantum_2007}.

The magnetic properties of SCBO have been extensively explored using Raman spectroscopy \cite{gozar_symmetry_2005, thirunavukkuarasu_magnetoelastic_2023}, neutron scattering \cite{aso_high_2005, gaulin_high-resolution_2004, fogh_field-induced_2024, fogh_spin_2024}, magnetic susceptibility \cite{kageyama_anomalous_1999}, heat capacity \cite{guo_quantum_2020, guo_deconfined_2025, jorge_crystal_2005, jimenez_quantum_2021} and electron spin resonance \cite{zorko_x-band_2004}. These studies have unveiled a wide spectrum of low-energy spin excitations, encompassing localized triplons as well as singlet and triplet bound states \cite{mcclarty_topological_2017}, which can be precisely manipulated through external perturbations such as magnetic fields, pressure, temperature, or light excitation \cite{fogh_field-induced_2024, shi_discovery_2022, guo_deconfined_2025, giorgianni_ultrafast_2023, kageyama_direct_2000}.

On the other hand, the electronic structure and associated high-energy electronic excitations in SCBO remain largely unexplored, especially from an experimental standpoint. Yet a detailed characterization of the $d$--$d$ transition energies of the Cu$^{2+}$ ions, which carry magnetic moments, is necessary to quantify the effects of the crystal-field environment and the degree of hybridization between Cu $d$-orbitals and O $p$-orbitals. This information is in turn essential for accurately modeling the Cu-O-Cu superexchange interactions that govern the magnetic behavior in SCBO.
\begin{figure*}[t]
    \centering
    \includegraphics[width=1\linewidth]{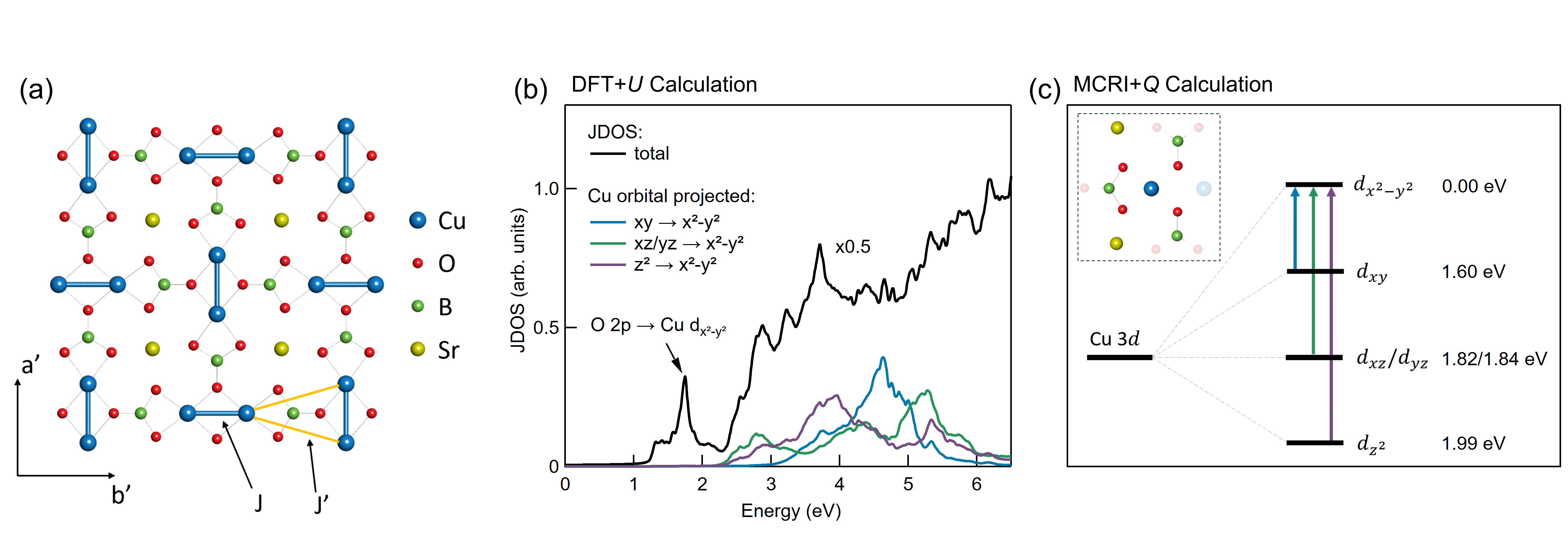}
    \caption{(a) Crystal structure of SrCu$_2$(BO$_3$)$_2$ viewed along the (001) direction. $a'$, $b'$ correspond to the crystallographic directions $[1,1,0]$ and $[\bar{1},1,0]$, respectively. Strontium (Sr) atoms are shown in yellow, oxygen (O) in red, boron (B) in green, and copper (Cu) in blue. (b) Total and projected JDOS of SCBO calculated using DFT+$U$ with $U = 4$~eV. The total JDOS is shown in black, while projections onto the Cu $3d$ orbitals are shown in blue, green, and violet. (c) Cu $d$‑orbital energy levels computed using the denoted as MRCI+$Q$ quantum chemistry method on an embedded 20-site cluster (dashed box) encompassing the CuO$_4$ plaquette, and including adjacent Cu, B, and Sr atoms.}
    \label{fig:structure}
\end{figure*}

Previous investigations of the electronic structure of SCBO have employed electron energy loss spectroscopy (EELS) at the O-$K$ edge combined with first-principles calculations based on density functional theory \cite{radtke_electronic_2008, radtke_momentum-resolved_2008}. These calculations, using the local density approximation (LDA)+$U$ where the Hubbard term $U$ is introduced to account for the effective on-site Coulomb interactions, qualitatively reproduced the observed EELS spectrum. However, the energy resolution in the EELS experiments was insufficient to properly resolve key electronic excitations, including charge-transfer (CT) transitions, or to access $d$–$d$ excitations, preventing full comparison with the computational model. Complementary investigations of the electronic structure have also been performed using optical spectroscopy, in which a pronounced absorption edge near 1.5 eV is attributed to Cu–O charge-transfer excitations. However, these measurements were made with a photon-energy bandwidth of only a few hundred meV, restricting the ability to fully characterize the charge-transfer peaks or resolve higher-energy electronic excitations~\cite{cherian_short-range_2014}. 

A comprehensive high-resolution experimental characterization of the electronic structure and the charge-transfer and intra-atomic $d$–$d$ transitions is still required to validate the computational models, especially those used to extract magnetic exchange interactions from band structure calculations. In particular, a detailed understanding of the electronic degrees of freedom is crucial to determine how they might underpin its exotic magnetic behavior. 




In this work, we combine Cu $L_3$-edge resonant inelastic x-ray scattering (RIXS) with broadband optical spectroscopy to determine the local and charge-transfer excitation spectrum of SCBO. The RIXS measurements directly probe the intra-atomic Cu $d$--$d$ excitations and reveal a well-defined manifold of localized crystal-field excitations in the 1.8--2.4~eV range. The extracted excitation energies and their polarization dependence are in good agreement with the Davidson-corrected internally contracted multireference configuration interaction \cite{werner_knowels_mrci_1988, langhoff_davidson_1974} (denoted as MRCI+$Q$) calculations, which explicitly account for strong local electronic correlations.

The optical measurements provide complementary access to the charge-transfer sector over an extended energy range. Infrared reflectivity and ellipsometry reveal a low-energy absorption onset near 1.2--1.6~eV together with a broader higher-energy structure centered around 4.5~eV. These features are assigned to O $2p \rightarrow$ Cu $3d$ charge-transfer excitations and are qualitatively reproduced by DFT+$U$ calculations. Importantly, no distinct $d$--$d$ features appear in the optical response, confirming their predominantly dipole-forbidden character and underscoring the complementarity of RIXS and optical spectroscopy.

By comparing DFT+$U$ and quantum-chemistry predictions with each other and with x-ray and optical spectroscopy experiments, we gauge the impact of strong local correlations and establish the hierarchy of local crystal-field and charge-transfer energy scales in SCBO. This combined approach benchmarks electronic-structure methods and provides experimentally constrained input for microscopic descriptions of Cu--O--Cu superexchange in this paradigmatic frustrated quantum magnet.

\section{Electronic Structure of $\rm{SrCu_2(BO_3)_2}$}
SCBO crystallizes in a tetragonal structure (space group $I\bar{4}2m$), in which Cu$^{2+}$ ions (each carrying spin $S=1/2$) form a planar pattern of orthogonal spin dimers embedded in the Cu–O–B networks within the $ab$ plane~\cite{vecchini_structural_2009}. As shown in Fig.~1(a), these dimers tile the plane in the characteristic geometry of the Shastry–Sutherland lattice—a frustrated arrangement in which strong intradimer exchange interactions $J$, oriented along the $a'$ and $b'$ crystallographic directions, alternate with weaker and frustrated interdimer couplings $J'$. This geometry gives rise to an exact dimer-singlet ground state and underpins hallmark phenomena in SCBO, including magnetization plateaus, bound triplon excitations, and other field-induced quantum phases \cite{kageyama_anomalous_1999, fogh_field-induced_2024, shi_discovery_2022}.

The magnetic interactions in SCBO are governed by Cu–O–Cu superexchange within the orthogonal dimer network. A dominant intradimer coupling $J$ competes with a weaker, frustrated interdimer interaction $J'$, while a small interlayer exchange $J''$ provides an additional perturbation~\cite{kodama_magnetic_2002}. These exchange pathways reflect the underlying electronic structure: hybridization between Cu $3d_{x^2-y^2}$ and O $2p$ orbitals mediates the antiferromagnetic superexchange,  while strong electronic correlations split the Cu-derived states into upper and lower Hubbard bands, with O $2p$ states lying energetically in between~\cite{radtke_electronic_2008}. Crystal-field effects lift the degeneracy of the Cu $3d^9$ orbitals, stabilizing the $d_{x^2-y^2}$ orbital as the highest-energy state hosting a single hole, while the remaining orbitals are fully occupied. In the electron picture, the $d$--$d$ excitations correspond to transitions from the $d_{xy}$, $d_{yz}$, $d_{xz}$, and $d_{z^2}$ orbitals to the partially occupied $d_{x^2-y^2}$ orbital. Additionally, the lowest-energy interband excitation is a charge-transfer transition from the O $2p$ states to the Cu $d_{x^2 - y^2}$ orbital, consistent with an insulating character and an optical gap of order 1–1.5 eV, as suggested by infrared spectroscopy~\cite{cherian_short-range_2014}.
\begin{figure*}[ht]
    \centering
    \includegraphics[width=\linewidth]{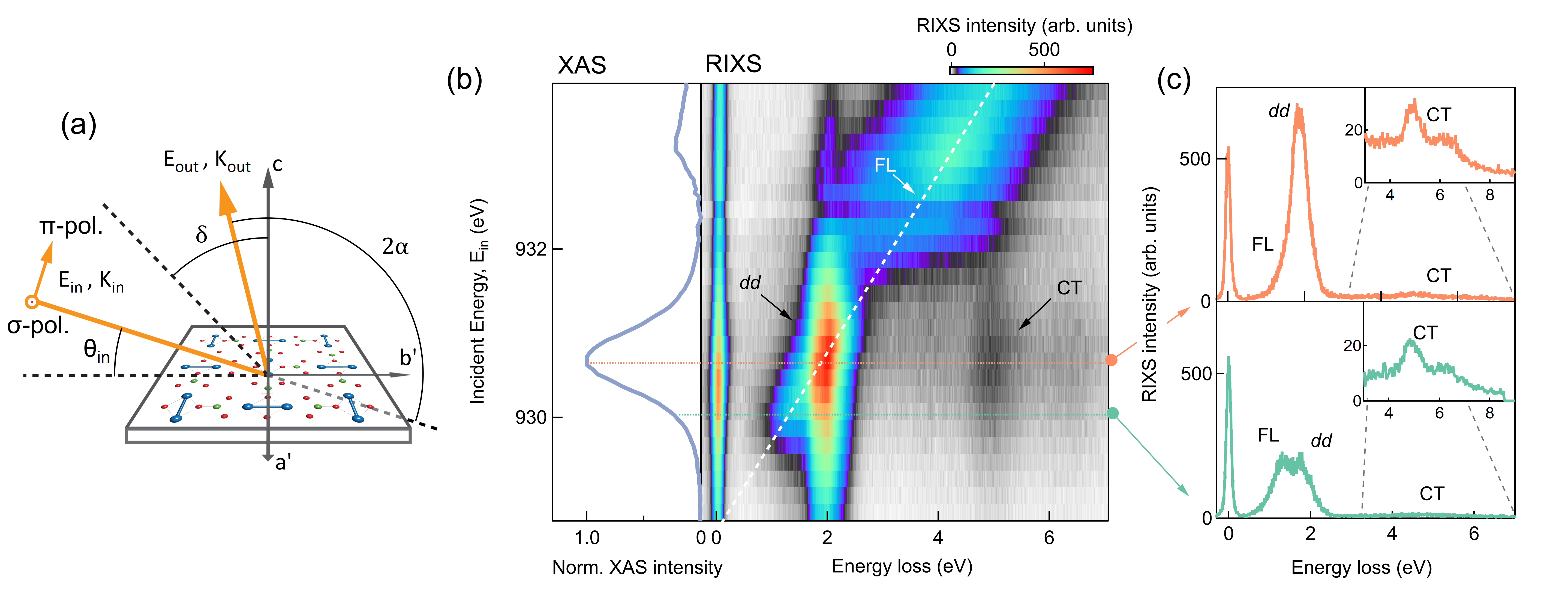}
\caption{(a) Schematic of the experimental geometry used in the resonant inelastic x-ray scattering (RIXS) experiment. The incident x-ray beam, with energy $E_{\text{in}}$ and wave-vector $k_{\text{in}}$ illuminates the sample at an angle $\theta_{in}$ relative to the sample surface. The scattered beam, with energy $E_{out}$ and wave-vector $k_{\text{out}}$ is collected by the RIXS spectrometer. The momentum transfer $\mathbf{q}$ forms an angle $\delta$ with respect to the normal of the scattering plane, which corresponds to the crystallographic $c$-axis. In the experiment, $\delta$ is controlled by rotating the sample plane around the crystal $a'$ axis. (b) X-ray absorption spectrum (XAS) near the Cu $L_3$-edge measured with $\sigma$-polarized x-ray light as a function of the incident photon energy $E_{\text{in}}$, along with the corresponding RIXS intensity map plotted as a function of energy loss ($E_{\text{in}} - E_{out}$). Several features can be identified in the RIXS map: charge-transfer excitations (CT), $d$--$d$ excitations (dd), and a constant-emission-energy fluorescence (FL) feature, highlighted with a white dotted line. (c) Representative RIXS spectra collected at $E_{\text{in}} = 930.7$~eV (on resonance with the Cu $L_3$-edge) and at $E_{\text{in}} = 930.1$~eV (off resonance). Insets: zoomed-in view highlighting the high-energy features of the spectra. All the measurements have been performed at 10 K.
}
\label{fig:Emap}
\end{figure*}

To investigate the electronic structure and $d$-orbital excitations in SCBO, DFT+$U$ calculations were first performed using the experimental crystal structure. These calculations provide a qualitative description of the electronic states relevant for CT and $d$–$d$ excitations; full computational details are given in Appendix~A.

From the DFT+$U$ electronic structure, the total and Cu $d$-orbital–resolved joint density of states (JDOS) can be computed to identify the dominant electronic transitions and characteristic energy scales, Fig.~1(b). Here, an effective Hubbard interaction term $U=4$~eV was adopted, following Ref.~\cite{radtke_electronic_2008}, as this value reproduces the experimentally determined superexchange interactions in SCBO ($J=75 K$ and $J'=45 K$). (Results for other $U$ values are discussed in the Sec. S3 of the Supplementary Information.)

The lowest-energy feature in the DFT+$U$ JDOS appears at $\sim$1.8~eV and corresponds to O~$2p \rightarrow$ Cu~$3d_{x^2-y^2}$ charge-transfer excitations, consistent with infrared spectroscopy measurements~\cite{cherian_short-range_2014}. At higher energies, the JDOS exhibits additional contributions arising from intra-atomic $d$–$d$ transitions from the occupied Cu $3d_{xy}$, $3d_{xz}$, $3d_{yz}$, and $3d_{z^2}$ orbitals into the partially occupied $3d_{x^2-y^2}$ state. However, within DFT+$U$, these $d$–$d$ excitations are placed in the 4–5~eV range, significantly higher than typical Cu $d$–$d$ excitation energies observed in copper oxides~\cite{moretti_sala_energy_2011}. 

To obtain a more accurate \textit{ab initio} description of the Cu $d$--$d$ excitation energies, we performed quantum-chemistry calculations using a cluster-in-embedding approach based on complete active space self-consistent field \cite{roos_casscf_1980, kreplin_casscf_2020} (CASSCF) theory combined with multireference configuration interaction \cite{werner_knowels_mrci_1988} (MRCI), as described in Appendix~A. The size-consistency Davidson correction \cite{langhoff_davidson_1974} ($+Q$) was also included. The resulting $d$--$d$ excitation energies are summarized in Fig.~1(c). As in other planar Cu-based compounds, the in-plane $d_{xy}$ orbital is predicted to lie at higher energy than the out-of-plane $d_{xz/yz}$ and $d_{z^2}$ orbitals~\cite{huang_cuprates_2011, Nikolay_nickelate_2020}. 

These results highlight the distinct strengths of the two theoretical approaches employed in this work. The DFT+$U$ calculations provide a useful band-structure-based description of the orbital character and overall energy scale of the charge-transfer excitations, but they do not quantitatively reproduce the localized Cu $d$--$d$ manifold. By contrast, the multireference embedded cluster treatment is essential for the local crystal-field excitations probed by Cu $L_3$-edge RIXS, for which MRCI+$Q$ captures both the level ordering and the relevant excitation energies.

\section{$\rm Cu$ L$_3$-edge Resonant Inelastic X-ray Scattering}

The symmetry and energy scales of the $d$--$d$ excitations in SCBO are investigated experimentally using high-resolution Cu $L_3$-edge RIXS. RIXS is particularly well suited to probing local $d$--$d$ excitations because of its element specificity, resonant enhancement, and sensitivity to crystal-field and multiplet-split states~\cite{mitrano_exploring_2024, ament_resonant_2011}.

High-resolution RIXS measurements were carried out at the Veritas beamline of the MAX IV Synchrotron, Sweden~\cite{veritas}. The experiments were performed on high-quality SCBO single crystals with a thickness of 300~$\mu$m and a c-cut (001) orientation, grown using the traveling solvent floating zone method~\cite{KAGEYAMA199965}. The crystallographic orientation and sample quality were confirmed by Laue diffraction.

The incident photon energy was scanned between 925 and 945~eV across the Cu $L_3$ absorption edge. The measurements were performed at a base temperature of 10~K using a continuous-flow liquid-helium cryostat. The overall energy resolution, estimated from the full width at half maximum of the elastic line measured on a carbon tape sample, was approximately 130~meV.

The sample was mounted with the crystallographic $c$ axis and the $b'$ axis in the scattering plane, as illustrated in Fig.~2(a). The scattering angle $2\alpha$ was fixed at $145^\circ$. The incident x-ray beam, characterized by photon energy $E_{\mathrm{in}}$ and wavevector $\mathbf{k}_{\mathrm{in}}$, impinges on the sample at an angle $\theta_{\mathrm{in}}$ relative to the sample surface within the scattering plane, while the scattered beam, with photon energy $E_{\mathrm{out}}$ and wavevector $\mathbf{k}_{\mathrm{out}}$, is collected by the spectrometer. The linear polarization of the incident beam was set to be perpendicular ($\sigma$ polarization) or parallel ($\pi$ polarization) to the scattering plane.

The scattering geometry was parameterized by the deviation angle $\delta$, defined as the angle between the sample surface normal (c axis) and the bisector of the scattering angle. The momentum transfer $\mathbf{q}=\mathbf{k}_{\mathrm{out}}-\mathbf{k}_{\mathrm{in}}$ lies within the scattering plane and varies as $|\mathbf{q}|\propto \sin\delta$. The dispersion of the RIXS excitations was probed by varying $\delta$, which is achieved by rotating the sample around the $a'$ axis and thereby adjusting the incident angle $\theta_{\mathrm{in}}$.

Prior to the RIXS measurements, the Cu L$_3$-edge was characterized using the x-ray absorption spectrum (XAS) of SCBO in total-electron-yield (TEY) mode. Fig. 2(b) presents the XAS spectrum recorded at normal incidence ($\alpha=90^\circ$) with $\sigma$-polarized light, with the incident excitation photon energy $E_{in}$ varied from 925\,eV to 945\,eV. A pronounced absorption peak is observed at 930.7\,eV, corresponding to the Cu L$_3$ white-line feature, consistent with previous observations in Cu$^{2+}$ oxide systems. A secondary, less intense peak at 933.2\,eV is also observed; this feature likely arises from ligand-field multiplet effects or minor contributions from Cu in alternative oxidation states~\cite{xu_copper_2013}.

The RIXS spectra as a function of the incident photon energy, acquired under the same experimental geometry as the XAS, are shown in Fig.~2(b). The RIXS map reveals four prominent features: (i) an elastic (or quasi-elastic) peak centered at zero energy loss; (ii) $d$–$d$ excitations near 1.8 eV, resonantly enhanced at the white line energy; (iii) a fluorescence feature, whose energy loss increases with incident photon energy, characteristic of a fluorescence process; and (iv) high-energy charge-transfer (CT) excitations appearing above 4\,eV, which are independent of the excitation energy. These features are isolated in Fig. 2(c), where representative RIXS-cuts at incident energies $E_{in}$ of 930.1 eV and 930.7 eV are plotted. The $d$–$d$ peak at $\sim$1.8 eV is comparable to previous RIXS studies on other copper oxides~\cite{moretti_sala_energy_2011, martinelli_collective_2024, fumagalli_polarization-resolved_2019}.
In Sec.~IV, we analyze the angular and polarization dependence of these features to assign the individual orbital excitations.

\section{Angular-Dependent Cross Sections of $d$–$d$ Excitations}
RIXS intensities are sensitive to the dipole-transition matrix elements, which reflect the orbital symmetry and orientation of the electronic states involved~\cite{ament_resonant_2011, moretti_sala_energy_2011, kotani_resonant_2005}. Consequently, measuring the intensities of the individual $d$--$d$ peaks as a function of incident angle, and thus of deviation angle $\delta$, enables an unambiguous assignment of the excited states. Fig. 3 shows RIXS spectra collected at various deviation angles $\delta$ for both $\sigma$ and $\pi$ polarizations of the incident x-ray beam.
Each spectrum was fitted using multiple Gaussian functions to extract peak energies and relative intensities. The fitting procedure shows that the transition energies—corresponding to the centers of the Gaussian peaks—remain essentially constant within $\pm$ 0.02 eV over the full range of scanned $\delta$ values and the associated momentum transfers. This indicates that the observed excitations are non-dispersive and hence localised. The extended dataset, including the individual fits and their Gaussian components, is provided in Sec. S2 of the Supplementary Information.

\begin{figure}
    \centering
    \includegraphics[width=\linewidth]{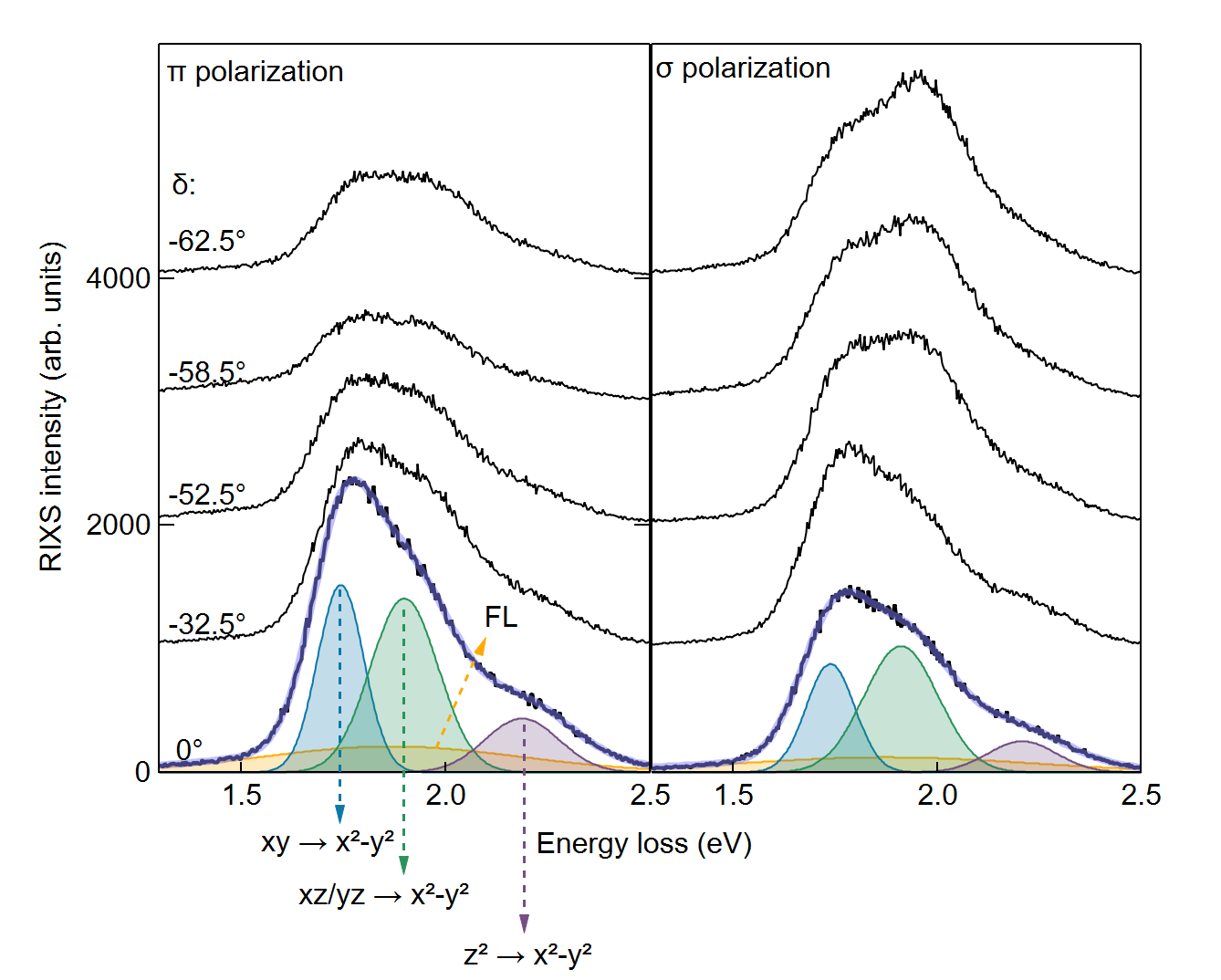}
    \caption{RIXS spectrum for $\pi$- and $\sigma$-polarized x-rays ($E_{in}=930.7$ eV) as a function of deviation angle $\delta$. The relative intensity of the individual $d$–$d$ excitations evolves as a function of $\delta$. These relative intensities were extracted through multi-component Gaussian fitting. Individual Gaussian lineshapes, corresponding to the $d$–$d$ excitations at $\delta = 0^\circ$, are shown as blue, green, and purple curves, while the total fit is shown as a violet line. The energies and symmetries of the $d$–$d$ excitations obtained are reported in Table I.}
    \label{fig:kdep}
\end{figure}

Table~I summarizes the experimental $d$--$d$ excitation energies and compares them with the values obtained from MRCI+$Q$ calculations. The overall agreement, within approximately 0.2~eV, strongly supports the assignment of these features to local crystal-field excitations of the Cu$^{2+}$ ion. The small systematic offset of the MRCI+$Q$ energies may originate from missing higher-order corrections, residual finite basis-set limitations, and from approximations inherent to the embedding scheme~\cite{mazurenko_first-principles_2008}.

Fig.~4 compares the experimental and MRCI+$Q$-based RIXS intensities \cite{bogdanov_rixs_2017} of the three resolved $d$--$d$ excitations as a function of deviation angle $\delta$ for both $\sigma$ and $\pi$ incident polarizations. The experimental intensities were extracted from fits to the spectra in Fig.~3 and corrected for self-absorption effects following Refs.~\cite{moretti_sala_energy_2011, zhang_unraveling_2022}; further details are provided in Sec.~S1 of the Supplementary Information.

As shown in Fig.~4, the calculated cross sections reproduce the main polarization selectivity and overall angular evolution of the experimental RIXS intensities for all three orbital channels. This agreement provides an additional validation of the orbital assignments derived from the excitation energies, since it links the wave-function character obtained from the multireference calculations to the measured scattering matrix elements. The remaining discrepancies are primarily in the absolute intensities and likely reflect the combined effects of background subtraction, self-absorption uncertainties, weak fluorescence contributions, and scattering channels not included in the present local model.

\setlength{\tabcolsep}{12pt}
\begin{table}[h]
\centering
\begin{tabular}{c c c }
\hline
$d$--$d$ Transition: & Exp. (eV) & MRCI+$Q$ (eV)\\
\hline
$d_{xy} \rightarrow d_{x^2-y^2}$ & 1.71 & 1.60 \\
$d_{yz}$/$d_{xz} \rightarrow d_{x^2-y^2}$ & 1.91 & 1.84/1.83 \\
$d_{z^2} \rightarrow d_{x^2-y^2}$ & 2.2 & 1.99 \\
\hline
\hline
\end{tabular}
\caption{Comparison of Cu $d$--$d$ excitation energies extracted from the experimental RIXS spectra with multireference quantum-chemistry results using MRCI+$Q$.}
\label{tab:dd-energies}
\end{table}

\begin{figure}
    \centering
    \includegraphics[width=\linewidth]{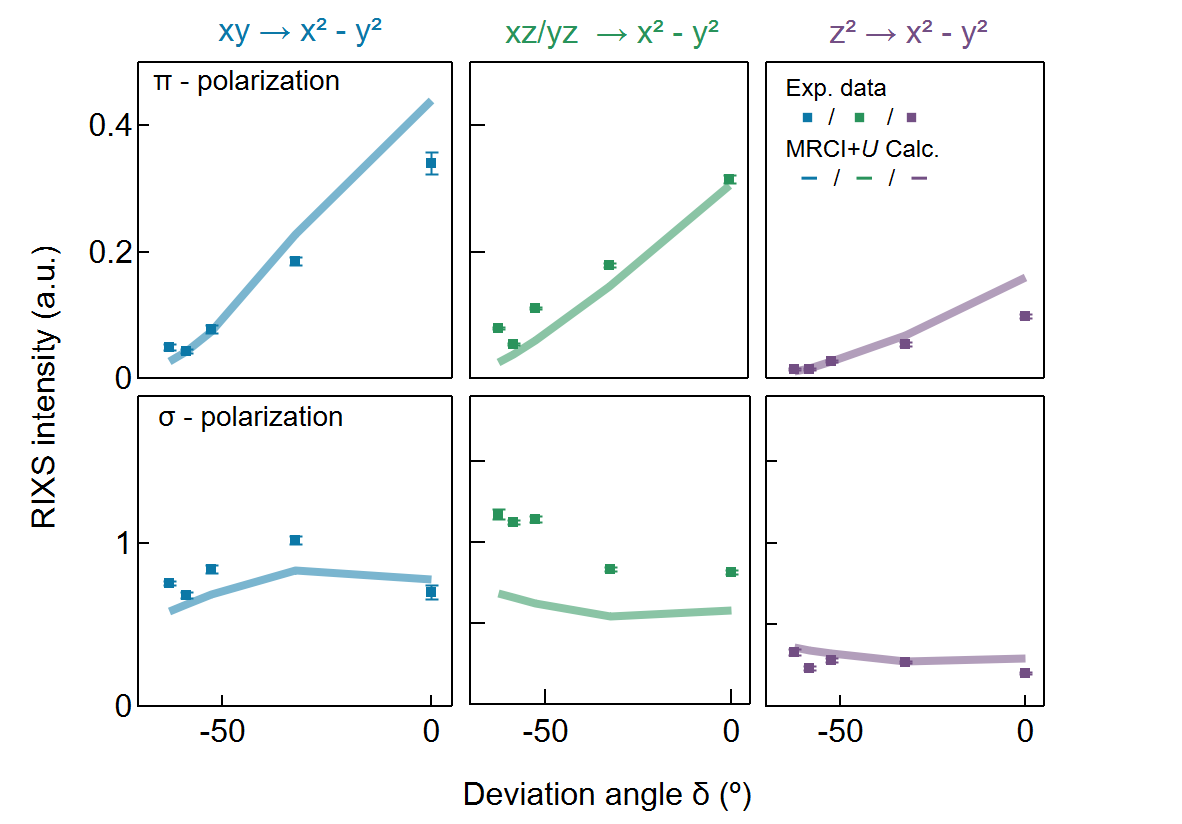}
    \caption{Intensity of the $d$--$d$ excitations as a function of deviation angle $\delta$, extracted from self-absorption–corrected experimental RIXS data and compared with relative intensities obtained from MRCI+$Q$ calculations (solid lines) for both $\sigma$- and $\pi$-polarized light. For each incident polarization, the experimental and calculated intensities were scaled using a single common normalization factor applied to all three orbital channels.}
    \label{fig:kdep_fits}
\end{figure}

\section{Broadband Optical Spectroscopy}
Having established the local $d$--$d$ excitation manifold with RIXS, we next turn to the optical response in order to probe the charge-transfer sector and to test whether any of the local crystal-field excitations acquire measurable dipole weight. Although $d$–$d$ transitions are typically dipole-forbidden and thus IR inactive, previous IR spectroscopy studies on layered copper oxides, such as CuGeO$_3$, have shown that lattice distortions allow these transitions to become optically active~\cite{bassi_optical_1996}. Therefore, broadband optical IR-spectroscopy is used to determine if a similar symmetry-breaking process in SCBO gives rise to a measurable dipole moment for $d$–$d$ excitations. Simultaneously, the optical spectroscopy is used to characterize the low-energy charge-transfer excitations, building a complete picture of the optical response.

Fourier-transform infrared (FTIR) spectroscopy in reflection geometry was performed on the same SCBO sample used in the RIXS measurements, using a commercial Bruker IFS66 spectrometer equipped with an optical cryostat for temperature control. The reflectivity was referenced against a platinum mirror to ensure accurate intensity calibration. Fig. 5(a) shows the reflectivity $R(\omega)$ from 1 eV to 2.5 eV at 4 K. A sharp drop in $R(\omega)$ at 1.6 eV marks the onset of optical absorption. A Kramers--Kronig-constrained variational fit was performed to extract the complex optical conductivity \(\tilde{\sigma}(\omega)\)~\cite{kuzmenko_kramerskronig_2005}. The fit to the reflectivity data is shown as a dashed line in Fig.~5(a). The resulting real part of the optical conductivity, \(\sigma_1(\omega)\), displayed in Fig.~5(b), exhibits a pronounced low-energy peak near 1.2~eV. No distinct feature is observed in the 1.5--1.7~eV range where the RIXS-measured $d$--$d$ excitations occur, confirming that these excitations carry at most a very weak dipole matrix element in the optical response.

To extend the spectral window, spectroscopic ellipsometry was used to determine the optical conductivity from 1.5 eV to 5.5 eV. These measurements were carried out at room temperature using a variable-angle Woollam ellipsometer. The resulting $\sigma_1(\omega)$ is presented in Fig.~5(b), and shows excellent agreement with the low photon energy response obtained from FTIR reflectivity. Notably, $\sigma_1(\omega)$ exhibits an additional broad feature centered around 4.5 eV, indicative of higher-energy interband transitions. The 4.5 eV peak is also weakly manifested in the Cu $L_3$ edge RIXS spectrum, as shown in Fig. 2(c). This consistency between RIXS and optical spectroscopy further supports the assignment of the high-energy feature to charge-transfer excitations.

\begin{figure}[t]
    \centering
    \includegraphics[width=\linewidth]{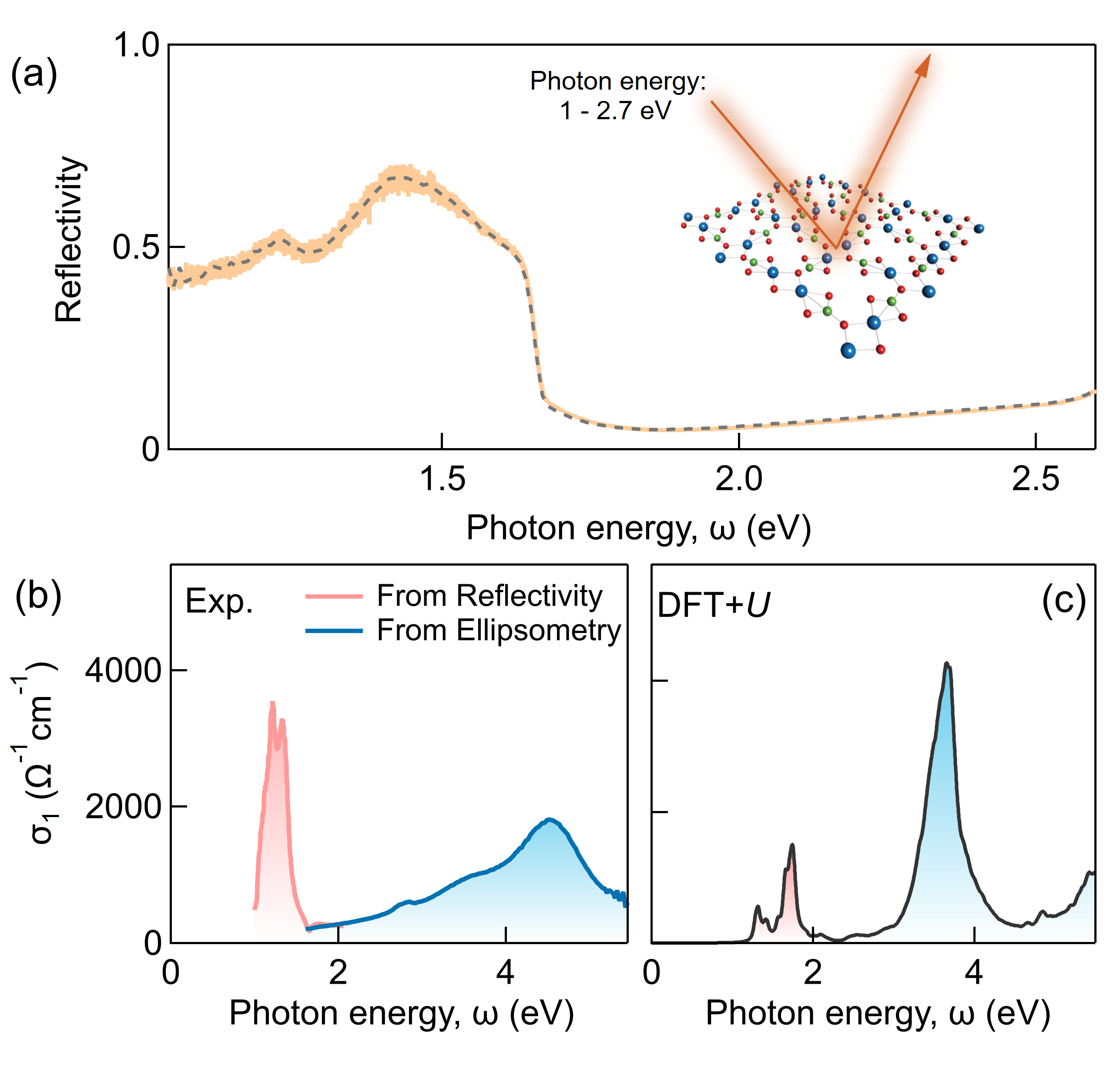}
    \caption{(a) FTIR reflectivity spectrum of SrCu$_2$(BO$_3$)$_2$ at 4~K (solid line) with Kramers--Kronig--constrained variational fit (dashed line). (b) Real part of the optical conductivity, $\sigma_1(\omega)$, extracted from the 4~K reflectivity fit (red) and extended to higher energies using room-temperature ellipsometry data (blue). Two distinct peaks in $\sigma_1(\omega)$ are associated with charge-transfer excitations involving O~2$p$ and Cu~$d$ states, as described in the text. (c) Calculated $\sigma_1(\omega)$ from DFT+$U$ band-structure calculations, depicting the theoretical optical response for comparison with experiment. }
    \label{fig:Ref}
\end{figure}

For comparison, the in-plane optical conductivity $\sigma_1(\omega)$ was calculated from the DFT+$U$ electronic structure with $U=4$~eV using the \texttt{epsilon.x} package of Quantum ESPRESSO~\cite{giannozzi_advanced_2017}. This code employs the independent particle approximation (IPA) to calculate the complex dielectric tensor from the DFT eigenvalues and eigenvectors. The resulting $\sigma_1(\omega)$, shown in Fig.~5(c), exhibits two pronounced peaks near 1.6\,eV and 3.6\,eV. It captures the overall structure and orbital character of the experimental optical response in Fig.~5(b), but exhibits noticeable quantitative deviations in the excitation energies, reflecting limitations of the independent-particle approximation and the static treatment of correlations. Based on the orbital-projected DOS (see Sec.~S3 of the Supplementary Information), both peaks originate from charge-transfer transitions involving the occupied O $2p$ orbitals and the unoccupied Cu $d_{x^{2}-y^{2}}$ orbital. The lower-energy peak arises from O $2p$ states near the top of the valence band, whereas the higher-energy peak is associated with transitions from O $2p$ states at larger binding energies. Contributions from other atomic orbitals are negligible, as their corresponding bands lie several electronvolts below the Fermi level and therefore do not participate in these optical excitations. For completeness, the total and orbital-projected DOS, together with the corresponding $\sigma_1(\omega)$ calculated for $U$ values ranging from 0 to 10\,eV, are presented in Sec.~S3 of the Supplementary Information.

\vspace{1px}

\section{Conclusions}
In summary, we have investigated the high-energy electronic excitations of SrCu$_2$(BO$_3$)$_2$ by combining Cu $L_3$-edge RIXS, broadband optical spectroscopy, and complementary electronic-structure calculations. The RIXS measurements resolve a localized manifold of Cu $d$--$d$ excitations in the 1.8--2.4~eV range. Their energies, orbital ordering, and polarization-dependent cross sections are consistently described by multireference quantum-chemistry calculations, supporting their assignment to local crystal-field excitations of the Cu$^{2+}$ ion.

Broadband optical spectroscopy reveals instead the charge-transfer excitations, with a low-energy absorption onset in the 1.2--1.6~eV range and a broader higher-energy structure extending to \(\sim 4.5\)~eV. These features are qualitatively reproduced by DFT+\(U\) calculations and are attributed predominantly to O \(2p\) \(\rightarrow\) Cu \(d\) interband excitations. The absence of corresponding $d$--$d$ features in the optical conductivity further confirms that these excitations carry negligible dipole spectral weight within experimental sensitivity, consistent with their predominantly dipole-forbidden character.

These results establish the hierarchy of local crystal-field and charge-transfer energy scales in SCBO and provide experimentally constrained input for microscopic descriptions of Cu--O--Cu superexchange. More broadly, this work defines a spectroscopic framework for connecting the high-energy electronic structure of the Shastry--Sutherland compound SrCu$_2$(BO$_3$)$_2$ to the low-energy magnetic Hamiltonians used to describe its frustration-driven quantum phases.

\section{Data Availability}
The experimental data supporting the findings of this article, along with the quantum chemistry results, are openly available~\cite{leinen_2026}. 
The density functional theory code and associated computational data are available through Materials Cloud~\cite{materialscloudProbingHighenergy}.

\section{Acknowledgments} 
This work was supported by the Swiss National Science Foundation (SNSF) under Grant No. 10001644 and SNSF Spark Grant No. 221173. N.C. acknowledges support from the NCCR MARVEL, a National Centre of Competence in Research, funded by the SNSF (Grant No. 205602), and from the Swiss National Supercomputing Centre (CSCS) for high-performance computing resources under the CSCS-PSI agreement. O.K.F is supported by the Swedish Research Council through Grant 2022-06217, the Foundation Blanceflor fellow scholarships for 2023 and 2024, and the Ruth and Nils-Erik Stenbäck Foundation.

\appendix
\section{Computational details}

\textbf{DFT+$U$ calculations.}

Density functional theory calculations within the DFT+$U$ framework were performed using the \textsc{Quantum ESPRESSO} package~\cite{giannozzi_advanced_2017}. A 44-atom unit cell with experimental structural parameters determined at 2~K~\cite{vecchini_structural_2009} was employed. Exchange–correlation effects were treated within the Perdew–Burke–Ernzerhof (PBE) generalized gradient approximation, augmented by an on-site effective Hubbard interaction term $U$ applied to the Cu $3d$ orbitals to account for strong electronic correlations. PBE ultrasoft projected-augmented-wave (PAW) pseudopotentials were used for all atomic species.

The magnetic structure was modeled in an antiferromagnetic configuration, with opposite spin polarization on the two Cu sites forming each dimer and a total magnetization constrained to zero. The Kohn–Sham wave functions were expanded in a plane-wave basis with kinetic-energy cutoffs of 90~Ry for the wave functions and 720~Ry for the charge density~\cite{prandini_precision_2018}. Brillouin-zone integrations for self-consistent calculations were performed using a $4 \times 4 \times 5$ Monkhorst–Pack $k$-point grid. For projected density-of-states calculations, a denser $k$-point mesh doubled with respect to the self-consistent grid was used to improve spectral resolution. Electronic occupations were treated using a smearing width of 0.015~Ry. All input files, output data, and post-processing scripts are available via the Materials Cloud archive~\cite{MC_archive}.

\vspace{8px}

\textbf{Quantum-chemistry calculations.}  
Multireference quantum-chemistry calculations were carried out using a cluster-in-embedding approach based on the same crystallographic structure. The quantum cluster comprises a single CuO$_4$ plaquette together with adjacent Cu, B, O, and Sr ions, as illustrated in the inset of Fig.~1(c). The cluster was embedded in an array of point charges constructed to reproduce the electrostatic field of the surrounding crystal using the extended Evjen scheme \cite{gelle_embedding_2008} with the help of the chargedel program \cite{bogdanov_chargedel}.

Electronic correlations within the cluster were treated by first performing the complete active space self-consistent field (CASSCF) calculations \cite{roos_casscf_1980, kreplin_casscf_2020} with the full Cu $d$-shell (9e, 5o) active space followed by multireference configuration interaction \cite{werner_knowels_mrci_1988} (MRCI) calculations to account for dynamical correlation effects by allowing single and double excitations from Cu 3d and O 2p orbitals. To account for higher-order quadruple excitations, the Davidson correction \cite{langhoff_davidson_1974} ($+Q$) was included, bringing the results into closer agreement with experiment (Table~II). In addition, calculations including spin-orbit coupling (SOC) were performed for both valence Cu $d^9$ and core-hole $2p^5 3d^{10}$ states \cite{schimmelpfennig_amfi_1998, berning_soc_2000}. The SOC vectors are reported in the Supplementary Information Sec.~S4. As shown in Table~II, inclusion of SOC introduces only a small correction to the MRCI+$Q$-calculated $d$--$d$ excitation energies. All quantum-chemistry calculations were performed using the \textsc{MOLPRO} software package~\cite{molpro_2020}.
For central Cu and O ions we employed cc-pwCVQZ-DK and cc-pVTZ-DK basis sets respectively; for adjacent Cu, O, B, Sr ions cc-pVDZ-DK basis set was used \cite{dunning_basis_1989, dejong_basis_2001, peterson_basis_2005, hill_basis_2017}

RIXS intensities at the MRCI+$Q$ level, including the angular dependence (Fig.~4), were computed using the procedure described in Ref.\,\onlinecite{bogdanov_rixs_2017}.

\begin{table*}[t]
\caption{Cu $d$--$d$ excitation energies obtained from the experimental RIXS spectra, compared with multireference quantum-chemistry calculations performed using MRCI, MRCI+$Q$, MRCI+SOC, and MRCI+$Q$+SOC levels, where $Q$ denotes the Davidson correction and SOC the spin-orbit coupling.}
\label{tab:dd-energies-app}
\begin{ruledtabular}
\begin{tabular*}{\textwidth}{@{\extracolsep{\fill}}lccccc}
$d$--$d$ transition & Exp. (eV) & MRCI & MRCI+$Q$ & MRCI+SOC & MRCI+$Q$+SOC \\
$d_{xy} \rightarrow d_{x^2-y^2}$ & 1.71 & 1.46 & 1.60 & 1.45 & 1.59 \\
$d_{yz}$/$d_{xz} \rightarrow d_{x^2-y^2}$ & 1.91 & 1.73/1.72 & 1.84/1.82 & 1.72/1.69 & 1.83/1.81 \\
$d_{z^2} \rightarrow d_{x^2-y^2}$ & 2.20 & 1.86 & 1.99 & 1.95 & 2.07 \\
\end{tabular*}
\end{ruledtabular}
\end{table*}

\FloatBarrier
\bibliographystyle{ieeetr}
\bibliography{SCBO_bibliography}

@article{kageyama_exact_1999,
	title = {Exact Dimer Ground State and Quantized Magnetization Plateaus in the Two-Dimensional Spin System  {SrCu\(_{2}\)(BO\(_{3}\))\(_{2}\)}},
	volume = {82},
	language = {en},
	number = {15},
	journal = {Phys. Rev. Lett.},
	author = {Kageyama, H and Yoshimura, K and Stern, R and Mushnikov, N V and Onizuka, K and Kato, M and Kosuge, K and Slichter, C P and Goto, T and Ueda, Y},
	year = {1999},
}

@article{cherian_short-range_2014,
	title = {Short-range magnetic interactions and optical band-edge physics in  {SrCu\(_{2}\)(BO\(_{3}\))\(_{2}\)}},
	volume = {90},
	copyright = {http://link.aps.org/licenses/aps-default-license},
	issn = {1098-0121, 1550-235X},
	url = {https://link.aps.org/doi/10.1103/PhysRevB.90.014405},
	doi = {10.1103/PhysRevB.90.014405},
	language = {en},
	number = {1},
	urldate = {2025-01-13},
	journal = {Phys. Rev. B},
	author = {Cherian, Judy G. and Tokumoto, Takahisa D. and Zhou, Haidong and McGill, Stephen A.},
	month = jul,
	year = {2014},
	pages = {014405},
}

@article{kodama_magnetic_2002,
	title = {Magnetic {Superstructure} in the {Two}-{Dimensional} {Quantum} {Antiferromagnet}  {SrCu\(_{2}\)(BO\(_{3}\))\(_{2}\)}},
	volume = {298},
	issn = {0036-8075, 1095-9203},
	url = {https://www.science.org/doi/10.1126/science.1075045},
	doi = {10.1126/science.1075045},
	language = {en},
	number = {5592},
	urldate = {2025-01-13},
	journal = {Science},
	author = {Kodama, K. and Takigawa, M. and Horvatić, M. and Berthier, C. and Kageyama, H. and Ueda, Y. and Miyahara, S. and Becca, F. and Mila, F.},
	month = oct,
	year = {2002},
	pages = {395--399},
}

@article{mcclarty_topological_2017,
	title = {Topological triplon modes and bound states in a {Shastry}–{Sutherland} magnet},
	volume = {13},
	issn = {1745-2473, 1745-2481},
	url = {https://www.nature.com/articles/nphys4117},
	doi = {10.1038/nphys4117},
	language = {english},
	number = {8},
	urldate = {2025-01-13},
	journal = {Nature Phys},
	author = {McClarty, P. A. and Krüger, F. and Guidi, T. and Parker, S. F. and Refson, K. and Parker, A. W. and Prabhakaran, D. and Coldea, R.},
	month = aug,
	year = {2017},
	pages = {736--741},
}

@article{radtke_electronic_2008,
	title = {Electronic structure of the quasi-two-dimensional spin-gap system  {SrCu\(_{2}\)(BO\(_{3}\))\(_{2}\)}: {Experiment} and theory},
	volume = {77},
	copyright = {http://link.aps.org/licenses/aps-default-license},
	issn = {1098-0121, 1550-235X},
	shorttitle = {Electronic structure of the quasi-two-dimensional spin-gap system {Sr} {Cu} 2 ( {B} {O} 3 ) 2},
	url = {https://link.aps.org/doi/10.1103/PhysRevB.77.125130},
	doi = {10.1103/PhysRevB.77.125130},
	language = {en},
	number = {12},
	urldate = {2025-01-13},
	journal = {Phys. Rev. B},
	author = {Radtke, G. and Saúl, A. and Dabkowska, H. A. and Gaulin, B. D. and Botton, G. A.},
	month = mar,
	year = {2008},
	pages = {125130},
}

@article{radtke_momentum-resolved_2008,
	title = {Momentum-resolved energy loss near edge structure in  {SrCu\(_{2}\)(BO\(_{3}\))\(_{2}\)}},
	volume = {108},
	copyright = {https://www.elsevier.com/tdm/userlicense/1.0/},
	issn = {03043991},
	url = {https://linkinghub.elsevier.com/retrieve/pii/S0304399108000399},
	doi = {10.1016/j.ultramic.2008.02.006},
	language = {en},
	number = {9},
	urldate = {2025-01-13},
	journal = {Ultramicroscopy},
	author = {Radtke, G.},
	month = aug,
	year = {2008},
	pages = {893--900},
}

@article{zayed_4-spin_2017,
	title = {4-spin plaquette singlet state in the {Shastry}–{Sutherland} compound  {SrCu\(_{2}\)(BO\(_{3}\))\(_{2}\)}},
	volume = {13},
	issn = {1745-2473, 1745-2481},
	url = {https://www.nature.com/articles/nphys4190},
	doi = {10.1038/nphys4190},
	language = {en},
	number = {10},
	urldate = {2025-01-13},
	journal = {Nature Phys},
	author = {Zayed, M. E. and Rüegg, Ch. and Larrea J., J. and Läuchli, A. M. and Panagopoulos, C. and Saxena, S. S. and Ellerby, M. and McMorrow, D. F. and Strässle, Th. and Klotz, S. and Hamel, G. and Sadykov, R. A. and Pomjakushin, V. and Boehm, M. and Jiménez–Ruiz, M. and Schneidewind, A. and Pomjakushina, E. and Stingaciu, M. and Conder, K. and Rønnow, H. M.},
	month = oct,
	year = {2017},
	pages = {962--966},
}

@article{fogh_field-induced_2024,
	title = {Field-induced bound-state condensation and spin-nematic phase in  {SrCu\(_{2}\)(BO\(_{3}\))\(_{2}\)} revealed by neutron scattering up to 25.9 {T}},
	volume = {15},
	issn = {2041-1723},
	url = {https://www.nature.com/articles/s41467-023-44115-z},
	doi = {10.1038/s41467-023-44115-z},
	language = {en},
	number = {1},
	urldate = {2025-01-13},
	journal = {Nat Commun},
	author = {Fogh, Ellen and Nayak, Mithilesh and Prokhnenko, Oleksandr and Bartkowiak, Maciej and Munakata, Koji and Soh, Jian-Rui and Turrini, Alexandra A. and Zayed, Mohamed E. and Pomjakushina, Ekaterina and Kageyama, Hiroshi and Nojiri, Hiroyuki and Kakurai, Kazuhisa and Normand, Bruce and Mila, Frédéric and Rønnow, Henrik M.},
	month = jan,
	year = {2024},
	pages = {442},
}

@article{kageyama_direct_2000,
	title = {Direct {Evidence} for the {Localized} {Single}-{Triplet} {Excitations} and the {Dispersive} {Multitriplet} {Excitations} in  {SrCu\(_{2}\)(BO\(_{3}\))\(_{2}\)}},
	volume = {84},
	copyright = {http://link.aps.org/licenses/aps-default-license},
	issn = {0031-9007, 1079-7114},
	url = {https://link.aps.org/doi/10.1103/PhysRevLett.84.5876},
	doi = {10.1103/PhysRevLett.84.5876},
	language = {en},
	number = {25},
	urldate = {2025-01-13},
	journal = {Phys. Rev. Lett.},
	author = {Kageyama, H. and Nishi, M. and Aso, N. and Onizuka, K. and Yosihama, T. and Nukui, K. and Kodama, K. and Kakurai, K. and Ueda, Y.},
	month = jun,
	year = {2000},
	pages = {5876--5879},
}

@article{shi_discovery_2022,
	title = {Discovery of quantum phases in the {Shastry}-{Sutherland} compound  {SrCu\(_{2}\)(BO\(_{3}\))\(_{2}\)} under extreme conditions of field and pressure},
	volume = {13},
	issn = {2041-1723},
	url = {https://www.nature.com/articles/s41467-022-30036-w},
	doi = {10.1038/s41467-022-30036-w},
	language = {en},
	number = {1},
	urldate = {2025-01-14},
	journal = {Nat Commun},
	author = {Shi, Zhenzhong and Dissanayake, Sachith and Corboz, Philippe and Steinhardt, William and Graf, David and Silevitch, D. M. and Dabkowska, Hanna A. and Rosenbaum, T. F. and Mila, Frédéric and Haravifard, Sara},
	month = apr,
	year = {2022},
	pages = {2301},
}

@article{shastry_exact_1981,
	title = {Exact ground state of a quantum mechanical antiferromagnet},
	volume = {108},
	issn = {0378-4363},
	url = {https://www.sciencedirect.com/science/article/pii/037843638190838X},
	doi = {https://doi.org/10.1016/0378-4363(81)90838-X},
	number = {1},
	journal = {Physica B+C},
	author = {Shastry, B. Sriram and Sutherland, Bill},
	year = {1981},
	pages = {1069--1070},
}

@article{zorko_x-band_2004,
	title = {X-band {ESR} study of the {2D} spin-gap system  {SrCu\(_{2}\)(BO\(_{3}\))\(_{2}\)}},
	volume = {272-276},
	issn = {0304-8853},
	url = {https://www.sciencedirect.com/science/article/pii/S0304885303019887},
	doi = {https://doi.org/10.1016/j.jmmm.2003.12.336},
	language = {english},
	journal = {J. Magn. Magn. Mat.},
	author = {Zorko, Andrej and Arčon, Denis and Nuttall, Christopher J. and Lappas, Alexandros},
	year = {2004},
	pages = {E699--E701},
	anannote = {Proceedings of the International Conference on Magnetism (ICM 2003)},
}

@article{vecchini_structural_2009,
	title = {Structural distortions in the spin-gap regime of the quantum antiferromagnet  {SrCu\(_{2}\)(BO\(_{3}\))\(_{2}\)}},
	volume = {182},
	issn = {0022-4596},
	url = {https://www.sciencedirect.com/science/article/pii/S0022459609004526},
	doi = {https://doi.org/10.1016/j.jssc.2009.09.017},
	number = {12},
	journal = {J. Solid State Chem.},
	author = {Vecchini, C. and Adamopoulos, O. and Chapon, L. C. and Lappas, A. and Kageyama, H. and Ueda, Y. and Zorko, A.},
	year = {2009},
	pages = {3275--3281},
}

@article{bassi_optical_1996,
	title = {Optical absorption of {CuGeO\(_{3}\)}},
	volume = {54},
	url = {https://link.aps.org/doi/10.1103/PhysRevB.54.R11030},
	doi = {10.1103/PhysRevB.54.R11030},
	number = {16},
	urldate = {2025-06-30},
	journal = {Phys. Rev. B},
	author = {Bassi, M. and Camagni, P. and Rolli, R. and Samoggia, G. and Parmigiani, F. and Dhalenne, G. and Revcolevschi, A.},
	month = oct,
	year = {1996},
	pages = {R11030--R11033},
}

@article{ament_resonant_2011,
	title = {Resonant inelastic x-ray scattering studies of elementary excitations},
	volume = {83},
	url = {https://link.aps.org/doi/10.1103/RevModPhys.83.705},
	doi = {10.1103/RevModPhys.83.705},
	number = {2},
	urldate = {2025-06-30},
	journal = {Rev. Mod. Phys.},
	author = {Ament, Luuk J. P. and van Veenendaal, Michel and Devereaux, Thomas P. and Hill, John P. and van den Brink, Jeroen},
	month = jun,
	year = {2011},
	pages = {705--767},
}

@article{xu_copper_2013,
	title = {Copper {L}-edge spectra: multiplet vs. multiple scattering theory},
	volume = {430},
	issn = {1742-6596},
	shorttitle = {Copper {L}-edge spectra},
	url = {https://dx.doi.org/10.1088/1742-6596/430/1/012010},
	doi = {10.1088/1742-6596/430/1/012010},
	language = {en},
	number = {1},
	urldate = {2025-06-30},
	journal = {J. Phys.: Conf. Ser.},
	author = {Xu, W and Zhang, X-L and Guo, Z-Y and Si, C and Zhao, Y-D and Marcelli, A and Chen, D-L and Wu, Z-Y},
	month = apr,
	year = {2013},
	pages = {012010},
}

@article{thirunavukkuarasu_magnetoelastic_2023,
	title = {Magnetoelastic interactions in  {SrCu\(_{2}\)(BO\(_{3}\))\(_{2}\)} studied by {Raman} scattering experiments and first principles calculations},
	volume = {107},
	url = {https://link.aps.org/doi/10.1103/PhysRevB.107.064410},
	doi = {10.1103/PhysRevB.107.064410},
	number = {6},
	urldate = {2025-07-08},
	journal = {Phys. Rev. B},
	author = {Thirunavukkuarasu, K. and Radtke, G. and Lu, Z. and Lazzeri, M. and Christianen, P. C. M. and Ballottin, M. V. and Dabkowska, H. A. and Gaulin, B. D. and Smirnov, D. and Jaime, M. and Saúl, A.},
	month = feb,
	year = {2023},
	pages = {064410},
}

@article{nomura_unveiling_2023,
	title = {Unveiling new quantum phases in the {Shastry}-{Sutherland} compound {SrCu\(_{2}\)(BO\(_{3}\))\(_{2}\)} up to the saturation magnetic field},
	volume = {14},
	copyright = {2023 The Author(s)},
	issn = {2041-1723},
	url = {https://www.nature.com/articles/s41467-023-39502-5},
	doi = {10.1038/s41467-023-39502-5},
	language = {en},
	number = {1},
	urldate = {2025-07-08},
	journal = {Nat Commun},
	author = {Nomura, T. and Corboz, P. and Miyata, A. and Zherlitsyn, S. and Ishii, Y. and Kohama, Y. and Matsuda, Y. H. and Ikeda, A. and Zhong, C. and Kageyama, H. and Mila, F.},
	month = jun,
	year = {2023},
	pages = {3769},
}

@article{miyahara_theory_2003,
	title = {Theory of the orthogonal dimer {Heisenberg} spin model for  {SrCu\(_{2}\)(BO\(_{3}\))\(_{2}\)}},
	volume = {15},
	issn = {0953-8984},
	url = {https://dx.doi.org/10.1088/0953-8984/15/9/201},
	doi = {10.1088/0953-8984/15/9/201},
	language = {en},
	number = {9},
	urldate = {2025-07-08},
	journal = {J. Phys.: Condens. Matter},
	author = {Miyahara, Shin and Ueda, Kazuo},
	month = feb,
	year = {2003},
	pages = {R327},
}

@article{giorgianni_ultrafast_2023,
	title = {Ultrafast frustration breaking and magnetophononic driving of singlet excitations in a quantum magnet},
	volume = {107},
	url = {https://link.aps.org/doi/10.1103/PhysRevB.107.184440},
	doi = {10.1103/PhysRevB.107.184440},
	number = {18},
	urldate = {2025-07-08},
	journal = {Phys. Rev. B},
	author = {Giorgianni, F. and Wehinger, B. and Allenspach, S. and Colonna, N. and Vicario, C. and Puphal, P. and Pomjakushina, E. and Normand, B. and Rüegg, Ch.},
	month = may,
	year = {2023},
	pages = {184440},
}

@article{kageyama_anomalous_1999,
	title = {Anomalous {Magnetizations} in {Single} {Crystalline}  {SrCu\(_{2}\)(BO\(_{3}\))\(_{2}\)}},
	volume = {68},
	issn = {0031-9015},
	url = {https://journals.jps.jp/doi/10.1143/JPSJ.68.1821},
	doi = {10.1143/JPSJ.68.1821},
	number = {6},
	urldate = {2025-07-08},
	journal = {J. Phys. Soc. Jpn.},
	author = {Kageyama, Hiroshi and Onizuka, Kenzo and Yamauchi, Touru and Ueda, Yutaka and Hane, Shingo and Mitamura, Hiroyuki and Goto, Tsuneaki and Yoshimura, Kazuyoshi and Kosuge, Koji},
	month = jun,
	year = {1999},
	pages = {1821--1823},
}

@article{jorge_crystal_2005,
	title = {Crystal symmetry and high-magnetic-field specific heat of  {SrCu\(_{2}\)(BO\(_{3}\))\(_{2}\)}},
	volume = {71},
	url = {https://link.aps.org/doi/10.1103/PhysRevB.71.092403},
	doi = {10.1103/PhysRevB.71.092403},
	number = {9},
	urldate = {2025-07-08},
	journal = {Phys. Rev. B},
	author = {Jorge, G. A. and Stern, R. and Jaime, M. and Harrison, N. and Bonča, J. and El Shawish, S. and Batista, C. D. and Dabkowska, H. A. and Gaulin, B. D.},
	month = mar,
	year = {2005},
	pages = {092403},
}

@article{gozar_symmetry_2005,
	title = {Symmetry and light coupling to phononic and collective magnetic excitations in  {SrCu\(_{2}\)(BO\(_{3}\))\(_{2}\)}},
	volume = {72},
	url = {https://link.aps.org/doi/10.1103/PhysRevB.72.064405},
	doi = {10.1103/PhysRevB.72.064405},
	number = {6},
	urldate = {2025-07-08},
	journal = {Phys. Rev. B},
	author = {Gozar, A. and Dennis, B. S. and Kageyama, H. and Blumberg, G.},
	month = aug,
	year = {2005},
	pages = {064405},
}

@article{gaulin_high-resolution_2004,
	title = {High-{Resolution} {Study} of {Spin} {Excitations} in the {Singlet} {Ground} {State} of  {SrCu\(_{2}\)(BO\(_{3}\))\(_{2}\)}},
	volume = {93},
	url = {https://link.aps.org/doi/10.1103/PhysRevLett.93.267202},
	doi = {10.1103/PhysRevLett.93.267202},
	number = {26},
	urldate = {2025-07-08},
	journal = {Phys. Rev. Lett.},
	author = {Gaulin, B. D. and Lee, S. H. and Haravifard, S. and Castellan, J. P. and Berlinsky, A. J. and Dabkowska, H. A. and Qiu, Y. and Copley, J. R. D.},
	month = dec,
	year = {2004},
	pages = {267202},
}

@article{aso_high_2005,
	title = {High {Energy}-{Resolution} {Inelastic} {Neutron} {Scattering} {Experiments} on {Triplet} {Bound} {State} {Excitations} in  {SrCu\(_{2}\)(BO\(_{3}\))\(_{2}\)}},
	volume = {74},
	issn = {0031-9015},
	url = {https://journals.jps.jp/doi/10.1143/JPSJ.74.2189},
	doi = {10.1143/JPSJ.74.2189},
	number = {8},
	urldate = {2025-07-08},
	journal = {J. Phys. Soc. Jpn.},
	author = {Aso, Naofumi and Kageyama, Hiroshi and Nukui, Katsuyuki and Nishi, Masakazu and Kadowaki, Hiroaki and Ueda, Yutaka and Kakurai, Kazuhisa},
	month = aug,
	year = {2005},
	pages = {2189--2192},
}

@article{mazurenko_first-principles_2008,
	title = {First-principles investigation of symmetric and antisymmetric exchange interactions of  {SrCu\(_{2}\)(BO\(_{3}\))\(_{2}\)}},
	volume = {78},
	url = {https://link.aps.org/doi/10.1103/PhysRevB.78.195110},
	doi = {10.1103/PhysRevB.78.195110},
	number = {19},
	urldate = {2025-07-08},
	journal = {Phys. Rev. B},
	author = {Mazurenko, V. V. and Skornyakov, S. L. and Anisimov, V. I. and Mila, F.},
	month = nov,
	year = {2008},
	pages = {195110},
}

@article{fogh_spin_2024,
	title = {Spin {Waves} and {Three} {Dimensionality} in the {High}-{Pressure} {Antiferromagnetic} {Phase} of  {SrCu\(_{2}\)(BO\(_{3}\))\(_{2}\)}},
	volume = {133},
	url = {https://link.aps.org/doi/10.1103/PhysRevLett.133.246702},
	doi = {10.1103/PhysRevLett.133.246702},
	number = {24},
	urldate = {2025-07-08},
	journal = {Phys. Rev. Lett.},
	author = {Fogh, Ellen and Giriat, Gaétan and Zayed, Mohamed E. and Piovano, Andrea and Boehm, Martin and Steffens, Paul and Safiulina, Irina and Hansen, Ursula B. and Klotz, Stefan and Soh, Jian-Rui and Pomjakushina, Ekaterina and Mila, Frédéric and Normand, Bruce and Rønnow, Henrik M.},
	month = dec,
	year = {2024},
	pages = {246702},
}

@article{guo_deconfined_2025,
	title = {Deconfined quantum critical point lost in pressurized  {SrCu\(_{2}\)(BO\(_{3}\))\(_{2}\)}},
	volume = {8},
	copyright = {2025 The Author(s)},
	issn = {2399-3650},
	url = {https://www.nature.com/articles/s42005-025-01976-8},
	doi = {10.1038/s42005-025-01976-8},
	language = {en},
	number = {1},
	urldate = {2025-07-08},
	journal = {Commun Phys},
	author = {Guo, Jing and Wang, Pengyu and Huang, Cheng and Chen, Bin-Bin and Hong, Wenshan and Cai, Shu and Zhao, Jinyu and Han, Jinyu and Chen, Xintian and Zhou, Yazhou and Li, Shiliang and Wu, Qi and Meng, Zi Yang and Sun, Liling},
	month = feb,
	year = {2025},
	pages = {75},
}

@article{guo_quantum_2020,
	title = {Quantum {Phases} of  {SrCu\(_{2}\)(BO\(_{3}\))\(_{2}\)} from {High}-{Pressure} {Thermodynamics}},
	volume = {124},
	url = {https://link.aps.org/doi/10.1103/PhysRevLett.124.206602},
	doi = {10.1103/PhysRevLett.124.206602},
	number = {20},
	urldate = {2025-07-08},
	journal = {Phys. Rev. Lett.},
	author = {Guo, Jing and Sun, Guangyu and Zhao, Bowen and Wang, Ling and Hong, Wenshan and Sidorov, Vladimir A. and Ma, Nvsen and Wu, Qi and Li, Shiliang and Meng, Zi Yang and Sandvik, Anders W. and Sun, Liling},
	month = may,
	year = {2020},
	pages = {206602},
}

@article{moretti_sala_energy_2011,
	title = {Energy and symmetry of dd excitations in undoped layered cuprates measured by {Cu} {L\(_{3}\)} resonant inelastic x-ray scattering},
	volume = {13},
	url = {https://dx.doi.org/10.1088/1367-2630/13/4/043026},
	doi = {10.1088/1367-2630/13/4/043026},
	number = {4},
	journal = {New J. Phys.},
	author = {Moretti Sala, M and Bisogni, V and Aruta, C and Balestrino, G and Berger, H and Brookes, N B and Luca, G M de and Di Castro, D and Grioni, M and Guarise, M and Medaglia, P G and Miletto Granozio, F and Minola, M and Perna, P and Radovic, M and Salluzzo, M and Schmitt, T and Zhou, K J and Braicovich, L and Ghiringhelli, G},
	month = apr,
	year = {2011},
	pages = {043026},
}

@article{martinelli_collective_2024,
	title = {Collective {Nature} of {Orbital} {Excitations} in {Layered} {Cuprates} in the {Absence} of {Apical} {Oxygens}},
	volume = {132},
	url = {https://link.aps.org/doi/10.1103/PhysRevLett.132.066004},
	doi = {10.1103/PhysRevLett.132.066004},
	number = {6},
	urldate = {2025-07-08},
	journal = {Phys. Rev. Lett.},
	author = {Martinelli, Leonardo and Wohlfeld, Krzysztof and Pelliciari, Jonathan and Arpaia, Riccardo and Brookes, Nicholas B. and Di Castro, Daniele and Fernandez, Mirian G. and Kang, Mingu and Krockenberger, Yoshiharu and Kummer, Kurt and McNally, Daniel E. and Paris, Eugenio and Schmitt, Thorsten and Yamamoto, Hideki and Walters, Andrew and Zhou, Ke-Jin and Braicovich, Lucio and Comin, Riccardo and Sala, Marco Moretti and Devereaux, Thomas P. and Daghofer, Maria and Ghiringhelli, Giacomo},
	month = feb,
	year = {2024},
	pages = {066004},
}

@article{zhang_unraveling_2022,
	title = {Unraveling the nature of spin excitations disentangled from charge contributions in a doped cuprate superconductor},
	volume = {7},
	copyright = {2022 The Author(s)},
	issn = {2397-4648},
	url = {https://www.nature.com/articles/s41535-022-00528-5},
	doi = {10.1038/s41535-022-00528-5},
	language = {en},
	number = {1},
	urldate = {2025-07-09},
	journal = {npj Quantum Mater.},
	author = {Zhang, Wenliang and Agrapidis, Cliò Efthimia and Tseng, Yi and Asmara, Teguh Citra and Paris, Eugenio and Strocov, Vladimir N. and Giannini, Enrico and Nishimoto, Satoshi and Wohlfeld, Krzysztof and Schmitt, Thorsten},
	month = dec,
	year = {2022},
	pages = {123},
}

@article{fumagalli_polarization-resolved_2019,
	title = {Polarization-resolved {Cu L}\(_{3}\)-edge resonant inelastic x-ray scattering of orbital and spin excitations in {NdaB}\(_{2}\){Cu}\(_{3}\){O}\(_{7-\delta}\) },
	volume = {99},
	url = {https://link.aps.org/doi/10.1103/PhysRevB.99.134517},
	doi = {10.1103/PhysRevB.99.134517},
	number = {13},
	urldate = {2025-07-10},
	journal = {Phys. Rev. B},
	author = {Fumagalli, R. and Braicovich, L. and Minola, M. and Peng, Y. Y. and Kummer, K. and Betto, D. and Rossi, M. and Lefrançois, E. and Morawe, C. and Salluzzo, M. and Suzuki, H. and Yakhou, F. and Le Tacon, M. and Keimer, B. and Brookes, N. B. and Sala, M. Moretti and Ghiringhelli, G.},
	month = apr,
	year = {2019},
	pages = {134517},
}

@article{Nikolay_nickelate_2020,
  title = {Electronic correlations and magnetic interactions in infinite-layer ${\mathrm{NdNiO}}_{2}$},
  author = {Katukuri, Vamshi M. and Bogdanov, Nikolay A. and Weser, Oskar and van den Brink, Jeroen and Alavi, Ali},
  journal = {Phys. Rev. B},
  volume = {102},
  issue = {24},
  pages = {241112},
  numpages = {7},
  year = {2020},
  month = {Dec},
  publisher = {American Physical Society},
  doi = {10.1103/PhysRevB.102.241112},
  url = {https://link.aps.org/doi/10.1103/PhysRevB.102.241112}
}

@article{kuzmenko_kramerskronig_2005,
	title = {Kramers–{Kronig} constrained variational analysis of optical spectra},
	volume = {76},
	issn = {0034-6748},
	url = {https://doi.org/10.1063/1.1979470},
	doi = {10.1063/1.1979470},
	number = {8},
	urldate = {2025-07-14},
	journal = {Review of Scientific Instruments},
	author = {Kuzmenko, A. B.},
	month = jul,
	year = {2005},
	pages = {083108},
}

@article{giannozzi_advanced_2017,
	title = {Advanced capabilities for materials modelling with {Quantum} {ESPRESSO}},
	volume = {29},
	issn = {0953-8984},
	url = {https://dx.doi.org/10.1088/1361-648X/aa8f79},
	doi = {10.1088/1361-648X/aa8f79},
	language = {en},
	number = {46},
	urldate = {2025-07-15},
	journal = {J. Phys.: Condens. Matter},
	author = {Giannozzi, P and Andreussi, O and Brumme, T and Bunau, O and Buongiorno Nardelli, M and Calandra, M and Car, R and Cavazzoni, C and Ceresoli, D and Cococcioni, M and Colonna, N and Carnimeo, I and Dal Corso, A and de Gironcoli, S and Delugas, P and DiStasio, R A and Ferretti, A and Floris, A and Fratesi, G and Fugallo, G and Gebauer, R and Gerstmann, U and Giustino, F and Gorni, T and Jia, J and Kawamura, M and Ko, H-Y and Kokalj, A and Küçükbenli, E and Lazzeri, M and Marsili, M and Marzari, N and Mauri, F and Nguyen, N L and Nguyen, H-V and Otero-de-la-Roza, A and Paulatto, L and Poncé, S and Rocca, D and Sabatini, R and Santra, B and Schlipf, M and Seitsonen, A P and Smogunov, A and Timrov, I and Thonhauser, T and Umari, P and Vast, N and Wu, X and Baroni, S},
	month = oct,
	year = {2017},
	pages = {465901},
}

@article{Bose_spingaps_2005,
 ISSN = {00113891},
 URL = {http://www.jstor.org/stable/24110094},
 author = {Indrani Bose},
 journal = {Current Science},
 number = {1},
 pages = {62--70},
 publisher = {Temporary Publisher},
 title = {Spin gap antiferromagnets: materials and phenomena},
 urldate = {2025-07-17},
 volume = {88},
 year = {2005}
}

@article{balents_spin_2010,
	title = {Spin liquids in frustrated magnets},
	volume = {464},
	copyright = {2010 Springer Nature Limited},
	issn = {1476-4687},
	url = {https://www.nature.com/articles/nature08917},
	doi = {10.1038/nature08917},
	language = {en},
	number = {7286},
	urldate = {2025-07-17},
	journal = {Nature},
	author = {Balents, Leon},
	month = mar,
	year = {2010},
	pages = {199--208},
}

@article{ramirez_short-range_2025,
	title = {Short-range order and hidden energy scale in geometrically frustrated magnets},
	volume = {6},
	issn = {2633-5409},
	url = {https://pubs.rsc.org/en/content/articlelanding/2025/ma/d4ma00914b},
	doi = {10.1039/D4MA00914B},
	language = {en},
	number = {4},
	urldate = {2025-07-17},
	journal = {Mater. Adv.},
	author = {Ramirez, A. P. and Syzranov, S. V.},
	month = feb,
	year = {2025},
	pages = {1213--1229},
}

@article{lee_end_2008,
	title = {An {End} to the {Drought} of {Quantum} {Spin} {Liquids}},
	volume = {321},
	url = {https://www.science.org/doi/10.1126/science.1163196},
	doi = {10.1126/science.1163196},
	number = {5894},
	urldate = {2025-07-18},
	journal = {Science},
	author = {Lee, Patrick A.},
	month = sep,
	year = {2008},
	pages = {1306--1307},
}

@article{vasiliev_milestones_2018,
	title = {Milestones of low-{D} quantum magnetism},
	volume = {3},
	copyright = {2018 The Author(s)},
	issn = {2397-4648},
	url = {https://www.nature.com/articles/s41535-018-0090-7},
	doi = {10.1038/s41535-018-0090-7},
	language = {en},
	number = {1},
	urldate = {2025-07-18},
	journal = {npj Quant Mater},
	author = {Vasiliev, Alexander and Volkova, Olga and Zvereva, Elena and Markina, Maria},
	month = mar,
	year = {2018},
	pages = {18},
}

@article{prandini_precision_2018,
	title = {Precision and efficiency in solid-state pseudopotential calculations},
	volume = {4},
	copyright = {2018 The Author(s)},
	issn = {2057-3960},
	url = {https://www.nature.com/articles/s41524-018-0127-2},
	doi = {10.1038/s41524-018-0127-2},
	language = {en},
	number = {1},
	urldate = {2025-07-21},
	journal = {npj Comput Mater},
	author = {Prandini, Gianluca and Marrazzo, Antimo and Castelli, Ivano E. and Mounet, Nicolas and Marzari, Nicola},
	month = dec,
	year = {2018},
	note = {Publisher: Nature Publishing Group},
	pages = {72},
}

@article{mitrano_exploring_2024,
	title = {Exploring {Quantum} {Materials} with {Resonant} {Inelastic} {X}-{Ray} {Scattering}},
	volume = {14},
	url = {https://link.aps.org/doi/10.1103/PhysRevX.14.040501},
	doi = {10.1103/PhysRevX.14.040501},
	number = {4},
	urldate = {2025-07-22},
	journal = {Phys. Rev. X},
	author = {Mitrano, M. and Johnston, S. and Kim, Young-June and Dean, M. P. M.},
	month = dec,
	year = {2024},
	note = {Publisher: American Physical Society},
	pages = {040501},
}

@article{laflorencie_quantum_2007,
	title = {Quantum and {Thermal} {Transitions} {Out} of the {Supersolid} {Phase} of a {2D} {Quantum} {Antiferromagnet}},
	volume = {99},
	url = {https://link.aps.org/doi/10.1103/PhysRevLett.99.027202},
	doi = {10.1103/PhysRevLett.99.027202},
	number = {2},
	urldate = {2025-07-22},
	journal = {Phys. Rev. Lett.},
	author = {Laflorencie, Nicolas and Mila, Frédéric},
	month = jul,
	year = {2007},
	note = {Publisher: American Physical Society},
	pages = {027202},
}

@article{kotani_resonant_2005,
	title = {Resonant inelastic {X}-ray scattering in d and f electron systems},
	volume = {47},
	issn = {1434-6036},
	url = {https://doi.org/10.1140/epjb/e2005-00303-4},
	doi = {10.1140/epjb/e2005-00303-4},
	language = {en},
	number = {1},
	urldate = {2025-07-22},
	journal = {Eur. Phys. J. B},
	author = {Kotani, A.},
	month = sep,
	year = {2005},
	pages = {3--27},
}

@article{kang_resolving_2019,
	title = {Resolving the nature of electronic excitations in resonant inelastic x-ray scattering},
	volume = {99},
	url = {https://link.aps.org/doi/10.1103/PhysRevB.99.045105},
	doi = {10.1103/PhysRevB.99.045105},
	number = {4},
	urldate = {2025-09-02},
	journal = {Phys. Rev. B},
	author = {Kang, M. and Pelliciari, J. and Krockenberger, Y. and Li, J. and McNally, D. E. and Paris, E. and Liang, R. and Hardy, W. N. and Bonn, D. A. and Yamamoto, H. and Schmitt, T. and Comin, R.},
	month = jan,
	year = {2019},
	note = {Publisher: American Physical Society},
	pages = {045105},
}

@misc{veritas,
  author       = {{MAX IV Laboratory}},
  title        = {Veritas – RIXS Beamline at MAX IV},
  howpublished = {\url{https://www.maxiv.lu.se/beamlines-accelerators/beamlines/veritas/}},
  note         = {Accessed: 2025-10-29. Page last updated: September 5, 2025. Page manager: Conny Såthe.},
  year         = {2025}
}

@article{jimenez_quantum_2021,
	title = {A quantum magnetic analogue to the critical point of water},
	volume = {592},
	copyright = {2021 The Author(s), under exclusive licence to Springer Nature Limited},
	issn = {1476-4687},
	url = {https://www.nature.com/articles/s41586-021-03411-8},
	doi = {10.1038/s41586-021-03411-8},
	language = {en},
	number = {7854},
	urldate = {2025-10-21},
	journal = {Nature},
	author = {Jiménez, J. Larrea and Crone, S. P. G. and Fogh, E. and Zayed, M. E. and Lortz, R. and Pomjakushina, E. and Conder, K. and Läuchli, A. M. and Weber, L. and Wessel, S. and Honecker, A. and Normand, B. and Rüegg, Ch and Corboz, P. and Rønnow, H. M. and Mila, F.},
	month = apr,
	year = {2021},
	note = {Publisher: Nature Publishing Group},
	pages = {370--375},
}

@article{KAGEYAMA199965,
title = {Crystal growth of the two-dimensional spin gap system SrCu2(BO3)2},
journal = {Journal of Crystal Growth},
volume = {206},
number = {1},
pages = {65-67},
year = {1999},
issn = {0022-0248},
doi = {https://doi.org/10.1016/S0022-0248(99)00313-9},
url = {https://www.sciencedirect.com/science/article/pii/S0022024899003139},
author = {H Kageyama and K Onizuka and T Yamauchi and Y Ueda}
}

@dataset{MC_archive,
  author       = {Leinen, Tariq and Forslund, Ola K. And  Paris, Eugenio and Colonna, Nicola and Caputo, Marco and Nagamine, Gabriel and Puphal, Pascal and Teyssier, Jeremie and Sassa, Yasmine and Schmitt, Thorsten and Bogdanov, Nikolay A. and Daghofer, Maria and Cavalieri, Adrian and Giorgianni, Flavio},
  title        = {Probing high-energy electronic excitations in the Shastry-Sutherland compound SrCu$_2$(BO$_3$)$_2$ with resonant inelastic x-ray scattering and optical spectroscopy},
  month        = Dec,
  year         = 2025,
  publisher    = {Materials Cloud},
  doi          = {10.24435/materialscloud:9g-5a},
  url          = {https://doi.org/10.24435/materialscloud:9g-5a}
}

@article{molpro_2020,
    author = {Werner, Hans-Joachim and Knowles, Peter J. and Manby, Frederick R. and Black, Joshua A. and Doll, Klaus and Heßelmann, Andreas and Kats, Daniel and Köhn, Andreas and Korona, Tatiana and Kreplin, David A. and Ma, Qianli and Miller, Thomas F., III and Mitrushchenkov, Alexander and Peterson, Kirk A. and Polyak, Iakov and Rauhut, Guntram and Sibaev, Marat},
    title = {The Molpro quantum chemistry package},
    journal = {J. Chem. Phys.},
    volume = {152},
    number = {14},
    pages = {144107},
    year = {2020},
    month = {04},
    issn = {0021-9606},
    doi = {10.1063/5.0005081},
    url = {https://doi.org/10.1063/5.0005081},
}

@article{kreplin_casscf_2020,
    title={MCSCF optimization revisited. II. Combined first- and second-order orbital optimization for large molecules},
    volume={152},
    ISSN={1089-7690},
    url={http://dx.doi.org/10.1063/1.5142241},
    DOI={10.1063/1.5142241},
    number={7},
    journal={J. Chem. Phys.},
    publisher={AIP Publishing},
    author={Kreplin, David A. and Knowles, Peter J. and Werner, Hans-Joachim},
    year={2020},
    month=feb }

@article{langhoff_davidson_1974,
    title={Configuration interaction calculations on the nitrogen molecule},
	volume={8},
    ISSN={1097-461X},
    url={http://dx.doi.org/10.1002/qua.560080106},
    DOI={10.1002/qua.560080106},
    number={1},
    journal={Int. J. Quantum Chem.},
    publisher={Wiley},
    author={Langhoff, Stephen R. and Davidson, Ernest R.},
    year={1974},
    month=jan,
    pages={61–72}
    }

@article{bogdanov_rixs_2017,
    title={Orbital breathing effects in the computation of x-rayd-ion spectra in
                  solids byab initiowave-function-based methods},
    volume={29},
    ISSN={1361-648X},
    url={http://dx.doi.org/10.1088/1361-648X/29/3/035502},
    DOI={10.1088/1361-648x/29/3/035502},
    number={3},
    journal={J. Phys.: Condens. Matter},
    author={Bogdanov, Nikolay A and Bisogni, Valentina and Kraus, Roberto and
                  Monney, Claude and Zhou, Kejin and Schmitt, Thorsten and Geck,
                  Jochen and Mitrushchenkov, Alexander O and Stoll, Hermann and
                  van den Brink, Jeroen and Hozoi, Liviu},
    year={2016},
    pages={035502}
}

@article{huang_cuprates_2011,
  title = {Ab initio calculation of $d$-$d$ excitations in quasi-one-dimensional Cu ${d}^{9}$ correlated materials},
  author = {Huang, Hsiao-Yu and Bogdanov, Nikolay A. and Siurakshina, Liudmila and Fulde, Peter and van den Brink, Jeroen and Hozoi, Liviu},
  journal = {Phys. Rev. B},
  volume = {84},
  issue = {23},
  pages = {235125},
  numpages = {8},
  year = {2011},
  month = {Dec},
  publisher = {American Physical Society},
  doi = {10.1103/PhysRevB.84.235125},
  url = {https://link.aps.org/doi/10.1103/PhysRevB.84.235125}
}

@article{roos_casscf_1980,
    title={A complete active space SCF method (CASSCF) using a density matrix
           formulated super-CI approach},
    volume={48},
    ISSN={0301-0104},
    url={http://dx.doi.org/10.1016/0301-0104(80)80045-0},
    DOI={10.1016/0301-0104(80)80045-0},
    number={2},
    journal={Chem. Phys.},
    author={Roos, Björn O. and Taylor, Peter R. and Sigbahn, Per E.M.},
    year={1980},
    month=may,
    pages={157–173}
}

@article{werner_knowels_mrci_1988,
    author = {Werner, Hans‐Joachim and Knowles, Peter J.},
    title = {An efficient internally contracted multiconfiguration–reference configuration interaction method},
    journal = {J. Chem. Phys.},
    volume = {89},
    number = {9},
    pages = {5803-5814},
    year = {1988},
    month = {11},
    issn = {0021-9606},
    doi = {10.1063/1.455556},
    url = {https://doi.org/10.1063/1.455556},
}

@article{gelle_embedding_2008,
    title={Fast calculation of the electrostatic potential in ionic crystals by direct summation method},
    volume={128},
    ISSN={1089-7690},
    url={http://dx.doi.org/10.1063/1.2931458}, DOI={10.1063/1.2931458},
    number={24},
    journal={J. Chem. Phys.},
    publisher={AIP Publishing},
    author={Gellé, Alain and Lepetit, Marie-Bernadette},
    year={2008},
    month=jun
}

@software{bogdanov_chargedel,
  author       = {Bogdanov, Nikolay A.},
  title        = {chargedel},
  month        = jan,
  year         = 2021,
  note    = {{Z}enodo. https://doi.org/10.5281/zenodo.4444173},
  version      = {0.1},
  doi          = {10.5281/zenodo.4444173},
  url          = {https://doi.org/10.5281/zenodo.4444173}
}

@article{dunning_basis_1989,
    author = {Dunning, Thom H.},
    title = {Gaussian basis sets for use in correlated molecular calculations. I. The atoms boron through neon and hydrogen},
    journal = {J. Chem. Phys.},
    volume = {90},
    pages = {1007-1023},
    year = {1989},
    doi = {10.1063/1.456153}
}

@article{dejong_basis_2001,
    author = {de Jong, W. A and Harrison, R. J. and Dixon, D. A.},
    title = {Parallel Douglas–Kroll energy and gradients in NWChem: Estimating scalar relativistic effects using Douglas–Kroll contracted basis sets},
    journal = {J. Chem. Phys.},
    volume = {114},
    number = {1},
    pages = {48-53},
    year = {2001},
    month = {01},
    issn = {0021-9606},
    doi = {10.1063/1.1329891},
    url = {https://doi.org/10.1063/1.1329891},
}

@article{peterson_basis_2005,
    title={Systematically convergent basis sets for transition metals. II. Pseudopotential-based correlation consistent basis sets for the group 11 (Cu, Ag, Au) and 12 (Zn, Cd, Hg) elements},
    volume={114},
    ISSN={1432-2234},
    url={http://dx.doi.org/10.1007/s00214-005-0681-9}, DOI={10.1007/s00214-005-0681-9},
    number={4–5},
    journal={Theor. Chem. Acc.},
    publisher={Springer Science and Business Media LLC},
    author={Peterson, Kirk A and Puzzarini, Cristina}, year={2005},
    month=aug,
    pages={283–296}
}

@article{hill_basis_2017,
    author = {Hill, J. Grant and Peterson, Kirk A.},
    title = {Gaussian basis sets for use in correlated molecular calculations. XI. Pseudopotential-based and all-electron relativistic basis sets for alkali metal (K–Fr) and alkaline earth (Ca–Ra) elements},
    journal = {J. Chem. Phys.},
    volume = {147},
    number = {24},
    pages = {244106},
    year = {2017},
    month = {12},
    issn = {0021-9606},
    doi = {10.1063/1.5010587},
    url = {https://doi.org/10.1063/1.5010587},
}

@article{schimmelpfennig_amfi_1998,
    title={On the efficiency of an effective Hamiltonian in spin-orbit CI calculations},
    volume={286},
    ISSN={0009-2614},
    url={http://dx.doi.org/10.1016/s0009-2614(98)00120-1},
    DOI={10.1016/s0009-2614(98)00120-1},
    number={3–4},
    journal={Chem. Phys. Lett.},
    publisher={Elsevier BV},
    author={Schimmelpfennig, Bernd and Maron, Laurent and Wahlgren, Ulf and Teichteil, Christian and Fagerli, Hilde and Gropen, Odd},
    year={1998},
    month=apr,
    pages={261–266}
}

@article{berning_soc_2000,
    title={Spin-orbit matrix elements for internally contracted multireference configuration interaction wavefunctions},
    volume={98},
    ISSN={1362-3028},
    url={http://dx.doi.org/10.1080/00268970009483386},
    DOI={10.1080/00268970009483386},
    number={21},
    journal={Mol. Phys.},
    publisher={Informa UK Limited},
    author={Berning, Andreas and Schweizer, Marcus and Werner, Hans-Joachim and Knowles, Peter J. and Palmieri, Paolo},
    year={2000},
    month=nov,
    pages={1823–1833}
}

@misc{materialscloudProbingHighenergy,
	author = {},
	title = {{P}robing high-energy electronic excitations in the {S}hastry-{S}utherland compound {SrCu\(_{2}\)(BO\(_{3}\))\(_{2}\)} with resonant inelastic x-ray scattering and optical spectroscopy --- archive.materialscloud.org},
	howpublished = {\url{https://archive.materialscloud.org/record/2025.193}},
	year = {},
	note = {[Accessed 05-05-2026]},
}

@misc{leinen_2026, title={Electronic excitations in the Shastry-Sutherland compound {SrCu\(_{2}\)(BO\(_{3}\))\(_{2}\)} --- figshare.com}, url={https://figshare.com/articles/dataset/_/32209494/0}, DOI={10.6084/m9.figshare.32209494}, howpublished={DOI: 10.6084/m9.figshare.32209494}, abstractNote={<p dir="ltr">This dataset provides the experimental data and calculations associated with our spectroscopic study of the quantum magnet SrCu<sub>2</sub>(BO<sub>3</sub>)<sub>2</sub> (SCBO). While SCBO is a cornerstone material for studying the Shastry–Sutherland model and frustrated magnetism, this work focuses on characterizing the high-energy electronic structure, specifically the crystal-field split Cu 3d-orbitals, as well as charge-transfer excitations that underpin the magnetic structure of SCBO.</p>}, publisher={figshare}, author={}, year={2026}, month={May} }


\newpage





\end{document}


\title{\textbf{Supplemental Information:} Electronic excitations in the Shastry-Sutherland compound SrCu$_2$(BO$_3$)$_2$}

\author[1]{Tariq Leinen}
\affil[1]{Institute of Applied Physics, University of Bern, CH-3012 Bern, Switzerland}
\author[2, 3]{Ola K. Forslund}
\affil[2]{Institute of Physics, University of Zurich, CH-8057 Zürich, Switzerland}
\affil[3]{Department of Physics and Astronomy, Uppsala University, Box 516, SE-75120 Uppsala, Sweden}
\author[4]{Eugenio Paris}
\affil[4]{PSI Center for Photon Science, Paul Scherrer Institute, CH-5232 Villigen-PSI, Switzerland}%
\author[4]{Nicola Colonna}
\author[5]{Marco Caputo}
\affil[5]{MAX IV Laboratory, Lund University, SE-221 00 Lund, Sweden}
\author[2]{Johan Chang}
\author[1]{Gabriel Nagamine}
\author[5]{Takashi Tokushima}
\author[5]{Conny Såthe}
\author[6]{Pascal Puphal}
\affil[6]{Max Planck Institute for Solid State Research, Heisenbergstrasse 1, 70569 Stuttgart, Germany}%
\author[7]{Jeremie Teyssier}
\affil[7]{Department of Quantum Matter Physics, University of Geneva, 1211, Geneva, Switzerland}
\author[4]{Thorsten Schmitt}
\author[6]{Nikolay A. Bogdanov}
\author[8]{Maria Daghofer}
\affil[8]{Institute for Functional Matter and Quantum Technologies, Universität Stuttgart, 70550 Stuttgart, Germany}
\author[1, 4]{Adrian L. Cavalieri}
\author[1]{Flavio Giorgianni}

\maketitle

\section{Self-absorption correction}
The experimental RIXS spectral intensities as a function of the deviation angle $\delta$ were corrected by accounting for self-absorption effects. Varying the x-ray incidence angle changes the effective scattering path length within the sample, thereby modifying the absorption of incident photons and reabsorption of emitted photons. Accounting for this angle-dependent self-absorption is necessary to ensure that the extracted spectral intensities accurately represent the intrinsic RIXS response. In particular, the self-absorption correction has been used to compare the measured RIXS intensities of the individual $d$--$d$ transitions and the theoretical predictions (see Fig.~4 of the main text).

Self-absorption correction calculation procedure closely follows the methods described in Refs.~\cite{kang_resolving_2019, zhang_unraveling_2022}. The correction factors for the four RIXS scattering channels: $\sigma\rightarrow\sigma$, $\sigma\rightarrow\pi$, $\pi\rightarrow\pi$, and $\pi\rightarrow\sigma$, where $\sigma$ and $\pi$ denote the two orthogonal polarization states, are given by~\cite{kang_resolving_2019}:
\begin{equation} \label{eq:SI_correction_sigma}
    C_{\sigma \rightarrow (\sigma,\pi)}(\theta_{\mathrm{in}})
    = \mu_{\sigma}(E_{\mathrm{in}},\theta_{\mathrm{in}})
    + \mu_{\sigma,\pi}(E_{\mathrm{out}},\theta_{\mathrm{out}})
      \frac{-\hat{k}_{\mathrm{in}} \cdot \hat{n}}
           {\hat{k}_{\mathrm{out}} \cdot \hat{n}},
\end{equation}
and
\begin{equation} \label{eq:SI_correction_pi}
    C_{\pi \rightarrow (\pi,\sigma)}(\theta_{\mathrm{in}})
    = \mu_{\pi}(E_{\mathrm{in}},\theta_{\mathrm{in}})
    + \mu_{\sigma,\pi}(E_{\mathrm{out}},\theta_{\mathrm{out}})
      \frac{-\hat{k}_{\mathrm{in}} \cdot \hat{n}}
           {\hat{k}_{\mathrm{out}} \cdot \hat{n}}.
\end{equation}

Here, $\mu_{\sigma,\pi}(E_{\mathrm{in}},\theta_{\mathrm{in}})$ is the polarization-dependent absorption coefficient at the incident photon energy $E_{\mathrm{in}}$, where $\theta_{\mathrm{in}}$ is the angle between the incoming x-ray beam and the
sample surface (i.e., the grazing angle). Similarly,
$\mu_{\sigma,\pi}(E_{\mathrm{out}},\theta_{\mathrm{out}})$ is the absorption
coefficient at the emitted photon energy $E_{\mathrm{out}}$, and emission angle $\theta_{\mathrm{out}}$ relative to the sample
surface. The vectors $\hat{k}_{\mathrm{in}}$ and
$\hat{k}_{\mathrm{out}}$ denote the directions of the incoming and outgoing
photons, respectively, and $\hat{n}$ is the surface normal of the sample. The geometric
factor appearing in Eqs.~(S1)--(S2) is given by $\frac{\hat{k}_{\text{in}} \cdot \hat{n}}{\hat{k}_{\text{out}} \cdot \hat{n} } = \frac{\sin(\theta_{\text{in}})}{\sin(\theta_{\text{out}})}$.

The correction factors in Eqs. S1 and S2 can be calculated from the expression of the absorption coefficient at a generic photon energy $E$ and grazing angle
$\theta$~\cite{kang_resolving_2019}:
\begin{equation}
    \mu_{\sigma,\pi}(E, \theta)
    = \boldsymbol{\epsilon}_{\sigma,\pi}^{\mathrm{T}}
      \left[ R(\theta)^{\mathrm{T}}\, F(E)\, R(\theta) \right]
      \boldsymbol{\epsilon}_{\sigma,\pi},
\end{equation}
where $F(E)$ is the absorption tensor in the crystal reference frame. Because
SrCu$_2$(BO$_3)_2$ is tetragonal, $F(E)$ in the crystal coordinate system has the form:
\begin{equation}
F(E) = 
\begin{pmatrix}
    f_{\mathrm{a'a'}}(E) & 0 & 0 \\
    0 & f_{\mathrm{a'a'}}(E) & 0 \\
    0 & 0 & f_{\mathrm{cc}}(E)
\end{pmatrix}.
\end{equation}
The matrix $R(\theta)$ in Eq. S3 describes the rotation from the crystal frame to the
laboratory frame:
\begin{equation}
R(\theta) =
\begin{pmatrix}
    1 & 0           & 0 \\
    0 & \cos\theta  & \sin\theta \\
    0 & -\sin\theta & \cos\theta
\end{pmatrix},
\end{equation}
while $\epsilon_{\sigma} = (1,0,0)$ and $\epsilon_{\pi} =(0,0,1)$ are the polarization vectors.
Therefore, for the two polarization components, the absoption coefficients are:
\begin{equation}
    \mu_{\pi}(E, \theta) = \epsilon_{\pi}\left(R(\theta)^T  F(E) R(\theta) \right) \epsilon_{\pi} = \sin^2(\theta)f_\text{a'a'}(E) + \cos^2(\theta)f_\text{cc}(E)\text{,}
\end{equation}
and
\begin{equation}
    \mu_{\sigma}(E, \theta)  = \epsilon_{\sigma}\left(R(\theta)^T  F(E) R(\theta) \right) \epsilon_{\sigma}=  f_\text{a'a'}(E)\text{.}
\end{equation}
Substituting these absorption coefficients into the correction factors in Eqs. S1–S2 allows the self-absorption corrections to be written directly in terms of the elements of the scattering matrix, yielding:
\begin{align}
C_{\sigma \rightarrow \sigma}(\theta_{\mathrm{in}}) 
&= f_{\mathrm{a'a'}}(E_{\mathrm{in}})
   + f_{\mathrm{a'a'}}(E_{\mathrm{out}})
     \frac{\sin\theta_{\mathrm{in}}}{\sin\theta_{\mathrm{out}}}, 
\\[6pt]
C_{\sigma \rightarrow \pi}(\theta_{\mathrm{in}}) 
&= f_{\mathrm{a'a'}}(E_{\mathrm{in}})
   + \bigg[
       f_{\mathrm{a'a'}}(E_{\mathrm{out}})
       + \left( f_{\mathrm{cc}}(E_{\mathrm{out}}) 
              - f_{\mathrm{a'a'}}(E_{\mathrm{out}}) \right)
         \cos^{2}\theta_{\mathrm{out}}
     \bigg]
     \frac{\sin\theta_{\mathrm{in}}}{\sin\theta_{\mathrm{out}}},
\\[6pt]
C_{\pi \rightarrow \sigma}(\theta_{\mathrm{in}})
&= f_{\mathrm{a'a'}}(E_{\mathrm{in}})
   + \left( f_{\mathrm{cc}}(E_{\mathrm{in}}) 
          - f_{\mathrm{a'a'}}(E_{\mathrm{in}}) \right) 
     \cos^{2}\theta_{\mathrm{in}}
   + f_{\mathrm{a'a'}}(E_{\mathrm{out}})
     \frac{\sin\theta_{\mathrm{in}}}{\sin\theta_{\mathrm{out}}},
\\[6pt]
C_{\pi \rightarrow \pi}(\theta_{\mathrm{in}}) 
&= f_{\mathrm{a'a'}}(E_{\mathrm{in}})
   + \left( f_{\mathrm{cc}}(E_{\mathrm{in}}) 
          - f_{\mathrm{a'a'}}(E_{\mathrm{in}}) \right) 
     \cos^{2}\theta_{\mathrm{in}}  \nonumber\\
&\quad
   + \bigg[
       f_{\mathrm{a'a'}}(E_{\mathrm{out}})
       + \left( f_{\mathrm{cc}}(E_{\mathrm{out}}) 
              - f_{\mathrm{a'a'}}(E_{\mathrm{out}}) \right)
         \cos^{2}\theta_{\mathrm{out}}
     \bigg]
     \frac{\sin\theta_{\mathrm{in}}}{\sin\theta_{\mathrm{out}}}.
\end{align}

Experimentally, the individual components of $F(E)$ were obtained by measuring the XAS spectra in total electron yield (TEY) mode at different angles
$\theta_{\mathrm{in}}$. In general, the TEY signal at specific polarization is related to the absorption coefficient through~\cite{zhang_unraveling_2022}:
\begin{equation}
    \mathrm{TEY}_{\sigma,\pi}(E,\theta_{\mathrm{in}})
    \propto
    \mu_{\sigma,\pi}(E,\theta_{\mathrm{in}})\, L\,
    \frac{1}{
        \sin\theta_{\mathrm{in}}
        + \mu_{\sigma,\pi}(E,\theta_{\mathrm{in}})\, L
    },
\end{equation}
where $L$ is the effective electron escape depth. Because the photon
penetration depth $\lambda_{\sigma,\pi}(E,\theta_{\mathrm{in}})=\mu_{\sigma,\pi}(E,\theta_{\mathrm{in}})^{-1} $ is typically much larger than $L$ (i.e.~$\lambda_{\sigma,\pi}\gg L$), the
denominator is dominated by the geometrical factor $\sin\theta_{\mathrm{in}}$.
Under this experimentally well-satisfied condition, the TEY signal becomes
approximately proportional to the projected absorption coefficient:
\begin{equation}
    \mu_{\sigma,\pi}(E,\theta_{\mathrm{in}})
    \propto
    \mathrm{TEY}_{\sigma,\pi}(E,\theta_{\mathrm{in}})\,
    \sin\theta_{\mathrm{in}}.
\end{equation}

Thus, the components of the absorption tensor $F(E)$ can be obtained directly from TEY measurements performed at different $\theta_{\mathrm{in}}$. Specifically, the measurement of
$\mathrm{TEY}_{\sigma}(E,\theta_{\mathrm{in}})$ at normal incidence
$\theta_{\mathrm{in}} = 90^\circ$ allows us to extract $f_{\mathrm{a'a'}}$ using
Eq.~(S7), while a TEY measurement with $\pi$ polarization at
$\theta_{\mathrm{in}} = 10^\circ$ yields $f_{\mathrm{cc}}$ through
Eq.~(S6).

The experimentally determined $f_{\mathrm{a'a'}}$ and $f_{\mathrm{cc}}$
components are shown in Fig.~S1(a). In Fig.~S1(b), the corresponding correction factors 
$C_{\sigma \rightarrow (\sigma,\pi)}$ and 
$C_{\pi \rightarrow (\sigma,\pi)}$ were calculated using 
Eqs.~(S8)--(S12) as a function of the deviation angle $\delta$. 
The angle $\delta$ is related to the grazing angles 
$\theta_{\mathrm{in}}$ and $\theta_{\mathrm{out}}$ through $\delta = \theta_{\mathrm{in}} - \alpha = \theta_{\mathrm{out}} + \alpha - \pi$
where $\alpha = 72.5^\circ$ is the scattering angle determined by the
experimental scattering geometry.

As can be seen in Fig.~S1(b), \( C_{\sigma \rightarrow \sigma}(\delta) \simeq 
C_{\sigma \rightarrow \pi}(\delta) \) and 
\( C_{\pi \rightarrow \pi}(\delta) \simeq C_{\pi \rightarrow \sigma}(\delta) \).
This arises from the fact that, far from the Cu \(L_{3}\)-edge
resonance - specifically in the photon-energy range relevant for the RIXS
emitted photons (\(E_{\mathrm{out}} \approx 928\text{--}930~\mathrm{eV}\), see Fig.~S1(a) ) - the magnitudes of the absorption tensor components is comparable: $f_{\mathrm{cc}}(E_{\mathrm{out}}) \simeq f_{\mathrm{a'a'}}(E_{\mathrm{out}})$.
In this regime, the absorption anisotropy is small, resulting in nearly
identical self-absorption correction factors for the two outgoing polarization
channels ($\sigma$ and $\pi$). Consequently, the correction factor can be
taken as independent of the outgoing polarization state, which allows to
define: $C_{\sigma}(\delta) =
C_{\sigma \rightarrow (\pi,\sigma)}(\delta)$ and $C_{\pi}(\delta) =
C_{\pi \rightarrow (\pi,\sigma)}(\delta)$.

Therefore, the $\delta$-dependent self-absorption correction for the two
polarizations is given by
\begin{equation}
    I^{\mathrm{corr}}_{\sigma,\pi}(\delta)
    = C_{\sigma,\pi}(\delta)\,
      I^{\mathrm{meas}}_{\sigma,\pi}(\delta),
\end{equation}
where $I^{\mathrm{meas}}_{\sigma,\pi}(\delta)$ denotes the measured RIXS
intensity for incident $\sigma$- or $\pi$-polarized light as a function of the
deviation angle $\delta$. The correction factor $C_{\sigma,\pi}(\delta)$ was
applied to the experimental spectra, and the resulting corrected intensities
were compared with the theoretical predictions obtained using \textsc{edrixs}
(see main text and Fig.~4).
\begin{figure}[h!]
    \centering
    \includegraphics[width=\linewidth]{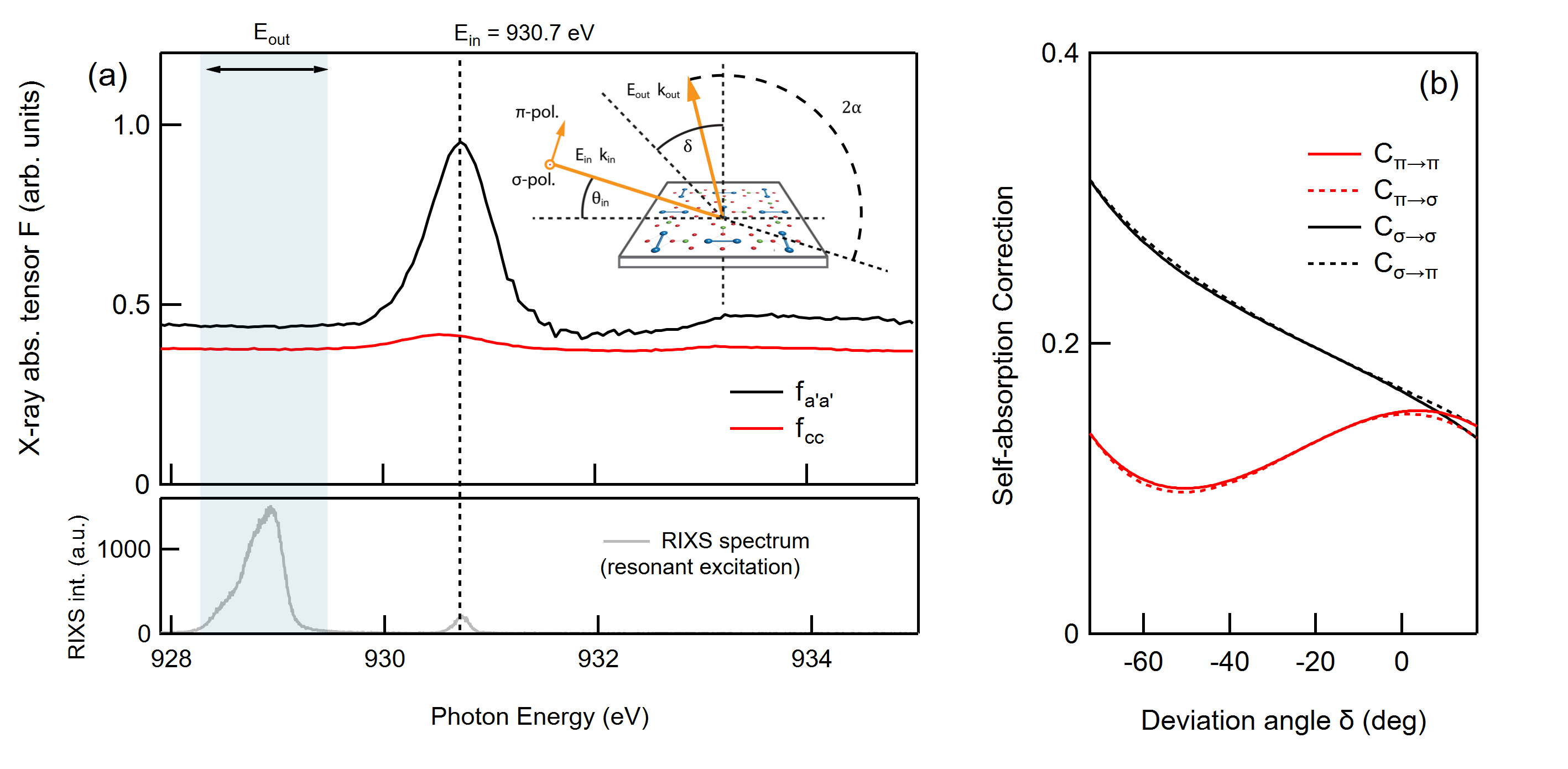}
\caption{(a) X-ray absorption tensor coefficients $f_{\mathrm{a'a'}}$ (black) and $f_{\mathrm{cc}}$ (red) shown in the upper panel, together with a representative RIXS spectrum at resonant excitation in the lower panel. The latter displays the inelastic response associated with $d$--$d$ excitations in a region where the XAS spectrum is relatively flat. At the photon energy relevant for the RIXS process, the absorption coefficients satisfy $f_{\mathrm{a'a'}} \simeq f_{\mathrm{cc}}$. The inset shows the RIXS scattering geometry, see Fig.~S1(a), as convetionally expressed in Ref.~\cite{moretti_sala_energy_2011}. 
(b) Self-absorption correction factors calculated from Eqs.~(S8)--(S11) as a function of $\delta$. One finds that $C_{\sigma \rightarrow \sigma}(\delta) \simeq C_{\sigma \rightarrow \pi}(\delta)$ and $C_{\pi \rightarrow \pi}(\delta) \simeq C_{\pi \rightarrow \sigma}(\delta)$, a direct consequence of the near equality $f_{\mathrm{a'a'}} \simeq f_{\mathrm{cc}}$ at the RIXS photon energy.}
    \label{fig:SI_Self_Abs}
\end{figure}

\section{RIXS intensity of $d$–$d$ transitions as a function of the deviation angle}

The intensities of the individual $d$–$d$ excitation peaks as a function of the deviation angle $\delta$, shown in Fig.~4 of the main text, were extracted from the experimentally measured RIXS spectra using multi-gaussian fitting. The RIXS spectra for both $\pi$- and $\sigma$-polarized incident x-ray light, along with their corresponding gaussian fit decompositions, are displayed in Fig.~S1. The elastic line was modeled with a gaussian centered at 0 eV, while the $d$–$d$ excitations were reproduced using three gaussian components. In addition, a weak and broad fluorescence background (discussed in the main text) was included using a broad gaussian function superimposed on the $d$–$d$ excitations. The extracted individual $d-d$ excitation peak intensities for $\pi$- and $\sigma$ polarization components were corrected by self-absorption corrector factor calculated from Eq. S14.
\begin{figure}[h!]
    \centering
    \includegraphics[width=0.65\linewidth]{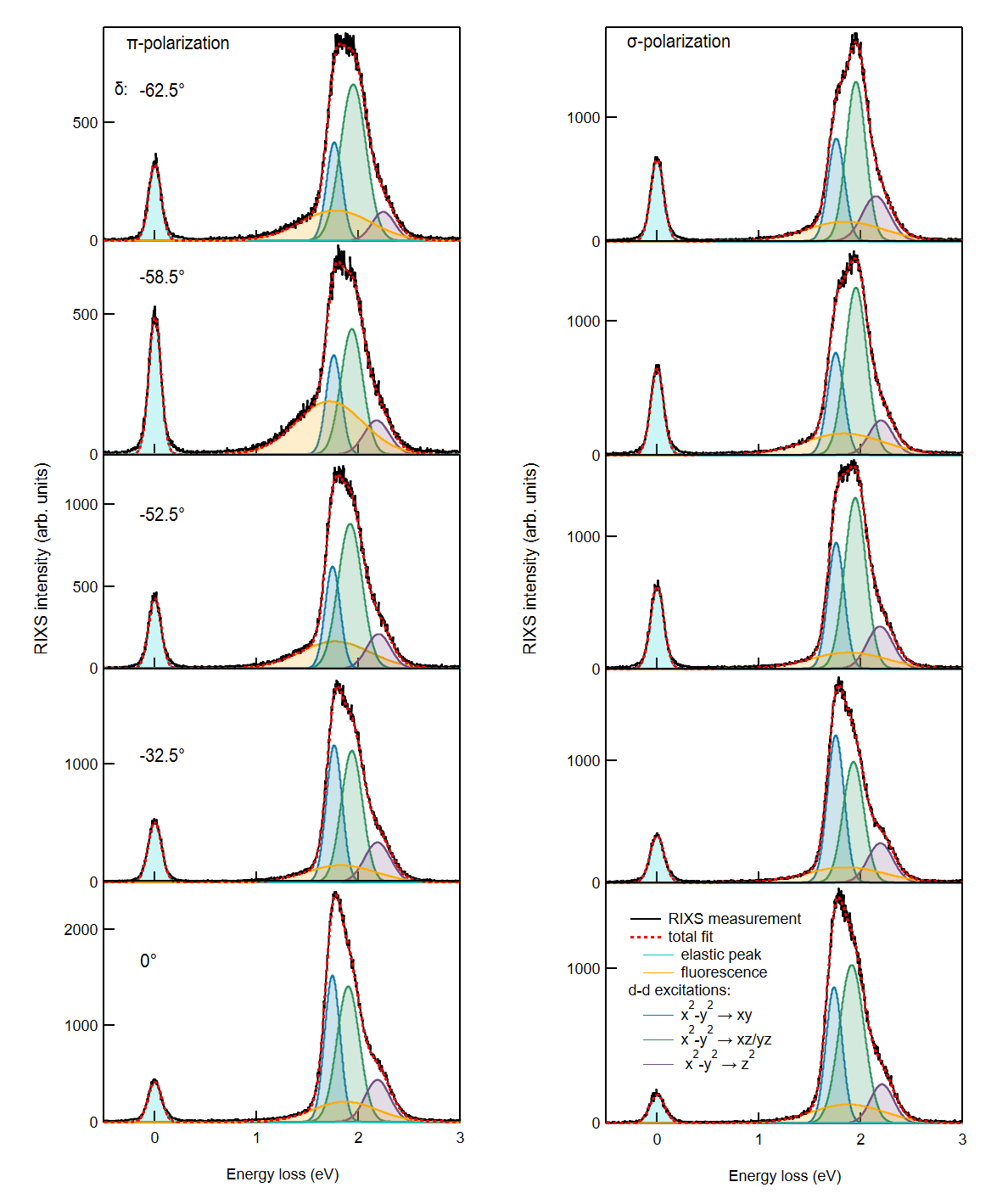}
    \caption{RIXS spectra as a function of the deviation angle $\delta$. The left column shows spectra measured with $\pi$-polarized, and the right column with $\sigma$-polarized incident x-ray beam. The evolution of the $d$–$d$ peak shape is captured by the relative intensities of the individual $d$–$d$ excitations obtained through multi-gaussian fitting, shown by the blue, green, and purple curves. The fluorescence background (yellow) and the elastic peak (light blue) were also fitted using gaussian functions.}
    \label{fig:SI_momentum_fits}
\end{figure}

\section{Assignment of peaks in the optical conductivity through DFT+U calculations}
From the real part of the optical conductivity measured by FTIR and ellipsometry, two prominent features are observed at approximately $1.2$~eV and $4.5$~eV. We assign these peaks on the basis of electronic band-structure calculations performed within the DFT+$U$ framework. As discussed in the main text, the inclusion of a Hubbard $U$ term is essential to capture the electronic correlations associated in SrCu$_2$(BO$_3$)$_2$. However, because the results of DFT+$U$ calculations depend on the chosen value of $U$, it is important to evaluate how different values of $U$ influence the resulting electronic band structure.
\begin{figure}[h!]
    \centering
    \includegraphics[width=0.66\linewidth]{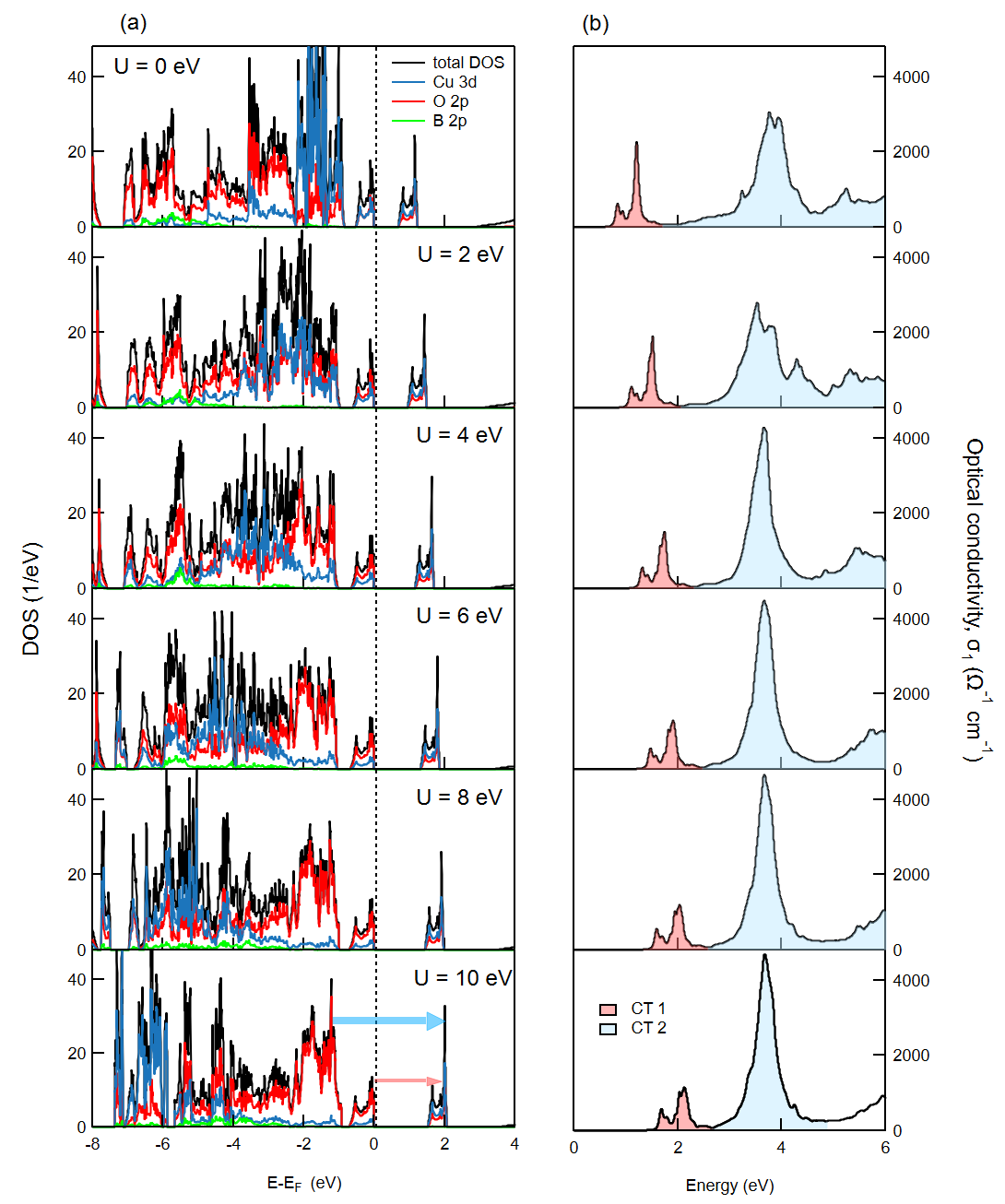}
\caption{ (a) DOS calculation by DFT+$U$ for $U = 0$, 2, 4, 6, 8, and 10~eV: total density of states (black) and orbital-projected contributions from Cu~$3d$ (blue), O~$2p$ (red), and B~$2p$ (green). Light blue and red arrows correspond to charge transfer excitations from O~$2p \rightarrow$ Cu~$3d_{x^{2}-y^{2}}$. (b) Corresponding calculated optical-conductivity spectra. The highlighted light blue and red features correspond to charge-transfer excitations involving O~$2p$ to Cu~$3d$ states. In the main text, the optical conductivity related to $U$ = 4 eV is reported, as this value reproduces the experimentally observed superexchange interactions.}
    \label{fig:SI_DFTpU}
\end{figure}
Fig.~S3(a) presents the total density of states (DOS) together with the orbital-resolved DOS of Cu~$3d$, O~$2p$, and B~$2p$ states for different values of $U$ from 0 to 10 eV. The corresponding calculated real-part optical-conductivity spectra are shown in Fig.~S3(b). Increasing $U$ shifts the Cu~$3d_{x^{2}-y^{2}}$ states to higher energies, while other $d$ orbitals orbitals shift downward; consequently, the separation between the O~$2p$ bands below the Fermi energy ($E_F$) and the Cu~$3d_{x^{2}-y^{2}}$ orbital increases systematically with $U$. This leads to an increase in the charge-transfer gap. Despite this shift, the overall DOS features remain largely unchanged as $U$ increases. Therefore, the two peaks observed experimentally in the optical conductivity at \(1.2\) and \(\sim 4.5\)~eV can be assigned to O~\(2p \rightarrow\) Cu~\(3d_{x^{2}-y^{2}}\) charge-transfer excitations, as indicated by the horizontal arrows in Fig.~S3(a).

Furthermore, the B~$2p$ states lie well below $-6$~eV relative to the Fermi level $E_F$. Due to this large energy separation, B~$2p$ cannot contribute to the optical feature observed near $4.5$~eV.

\section{Spin-orbit coupling correction to the MRCI excitation energies}
\label{sec:S4}

To estimate the effect of spin-orbit coupling (SOC) on the local Cu \(d\)-\(d\) excitations, the spin-free MRCI states were further coupled through the SOC Hamiltonian within the subspace of the low-energy valence Cu \(d^9\) configurations. In practice, a spin-orbit interaction matrix was constructed in the basis of the relevant spin-free MRCI states, and its diagonalization yielded the spin-orbit-coupled eigenstates and excitation energies.

The resulting SOC corrections are small, consistent with the predominantly crystal-field character of the excitations discussed in the main text. Table~S1 reports the decomposition of the spin-orbit-coupled states in the basis labeled by 1, 3, 5, 7, and 9. The states form the expected Kramers pairs, with \(S=1/2\) and opposite \(S_z=\pm 1/2\). For the lowest and highest states, the decomposition remains strongly dominated by a single spin-free basis state, whereas more sizable mixing is found for the intermediate levels. This indicates that SOC does not qualitatively alter the level scheme, but only introduces a modest redistribution of the wave-function character among nearby crystal-field states.

Overall, the SOC treatment confirms that the local excitation manifold is primarily governed by the crystal-field splitting, while spin-orbit effects provide only a small perturbative correction to the MRCI+\(Q\) energies.

\begin{table}[H]
\centering
\small
\setlength{\tabcolsep}{5pt}
\renewcommand{\arraystretch}{1.1}
\begin{tabular}{ccccccccc}
\hline
Nr & State & $S$ & $S_z$ & 1 & 3 & 5 & 7 & 9 \\
\hline
1  & 1 & 0.5 &  0.5 & 0.001\% & 0.005\% & 0.014\% & 0.157\% & 0.002\% \\
2  & 2 & 0.5 &  0.5 & 0.005\% & 2.733\% & 0.776\% & 5.460\% & 0.001\% \\
3  & 3 & 0.5 &  0.5 & 0.061\% & 5.649\% & 8.912\% & 9.900\% & 6.176\% \\
4  & 4 & 0.5 &  0.5 & 0.064\% & 4.411\% & 24.600\% & 17.777\% & 11.809\% \\
5  & 5 & 0.5 &  0.5 & 0.003\% & 0.079\% & 12.320\% & 3.070\% & 36.445\% \\
6  & 1 & 0.5 & -0.5 & 99.523\% & 0.189\% & 0.075\% & 0.033\% & 0.000\% \\
7  & 2 & 0.5 & -0.5 & 0.341\% & 86.605\% & 2.776\% & 1.294\% & 0.009\% \\
8  & 3 & 0.5 & -0.5 & 0.001\% & 0.208\% & 30.202\% & 30.960\% & 7.931\% \\
9  & 4 & 0.5 & -0.5 & 0.000\% & 0.114\% & 16.332\% & 20.868\% & 4.024\% \\
10 & 5 & 0.5 & -0.5 & 0.001\% & 0.006\% & 3.994\% & 10.480\% & 33.601\% \\
\hline
\end{tabular}
\caption{Decomposition of the spin-orbit-coupled states in the spin-free MRCI basis labeled by 1, 3, 5, 7, and 9. The calculated states form Kramers pairs with \(S=1/2\) and \(S_z=\pm 1/2\).}
\label{tab:state_decomposition}
\end{table}

\bibliographystyle{unsrt}
\bibliography{SCBO_bibliography}